\pdfoutput=1
\documentclass[twocolumn,english,amsmath,amssymb,aps,prl,reprint,floatfix,superscriptaddress,showpacs,booktabs]{revtex4}
\usepackage[latin9]{inputenc}
\usepackage{babel}
\usepackage{amstext}
\usepackage{graphicx}
\usepackage{esint}
\usepackage{subscript}
\usepackage[unicode=true,
 bookmarks=true,bookmarksnumbered=false,bookmarksopen=false,
 breaklinks=true,pdfborder={0 0 1},backref=false,colorlinks=false]
 {hyperref}
\hypersetup{pdftitle={Dynamical correlations and screened exchange on the experimental bench: spectral properties of the cobalt pnictide BaCo2As2},
 pdfauthor={A. van Roekeghem et. al.}}

\makeatletter

\providecommand{\tabularnewline}{\\}

\@ifundefined{textcolor}{}
{%
 \definecolor{BLACK}{gray}{0}
 \definecolor{WHITE}{gray}{1}
 \definecolor{RED}{rgb}{1,0,0}
 \definecolor{GREEN}{rgb}{0,1,0}
 \definecolor{BLUE}{rgb}{0,0,1}
 \definecolor{CYAN}{cmyk}{1,0,0,0}
 \definecolor{MAGENTA}{cmyk}{0,1,0,0}
 \definecolor{YELLOW}{cmyk}{0,0,1,0}
}

\@ifundefined{definecolor}{\usepackage{color}}{}

\makeatother

\begin{document}

\title{Dynamical correlations and screened exchange on the experimental
bench: spectral properties of the cobalt pnictide BaCo\textsubscript{2}As\textsubscript{2} }

\author{Ambroise van Roekeghem}

\email{vanroeke@cpht.polytechnique.fr}

\selectlanguage{english}%

\affiliation{Beijing National Laboratory for Condensed Matter Physics, and Institute
of Physics, Chinese Academy of Sciences, Beijing 100190, China}

\affiliation{Centre de Physique Théorique, Ecole Polytechnique, CNRS UMR 7644,
91128 Palaiseau, France}

\author{Thomas Ayral}

\affiliation{Centre de Physique Théorique, Ecole Polytechnique, CNRS UMR 7644,
91128 Palaiseau, France}

\affiliation{Institut de Physique Théorique (IPhT), CEA, CNRS, URA 2306, 91191
Gif-sur-Yvette, France }

\author{Jan M. Tomczak}

\affiliation{Institute of Solid State Physics, Vienna University of Technology,
A-1040 Vienna, Austria}

\author{Michele Casula}

\affiliation{CNRS and Institut de Minéralogie, de Physique des Matériaux et de
Cosmochimie, Université Pierre et Marie Curie, case 115, 4 place Jussieu,
FR-75252, Paris Cedex 05, France}

\author{Nan Xu}

\affiliation{Beijing National Laboratory for Condensed Matter Physics, and Institute
of Physics, Chinese Academy of Sciences, Beijing 100190, China}

\affiliation{Swiss Light Source, Paul Scherrer Institut, CH-5232 Villigen, Switzerland}

\author{Hong Ding}

\affiliation{Beijing National Laboratory for Condensed Matter Physics, and Institute
of Physics, Chinese Academy of Sciences, Beijing 100190, China}

\affiliation{Collaborative Innovation Center of Quantum Matter, Beijing, China}

\author{Michel Ferrero}

\affiliation{Centre de Physique Théorique, Ecole Polytechnique, CNRS UMR 7644,
91128 Palaiseau, France}

\author{Olivier Parcollet}

\affiliation{Institut de Physique Théorique (IPhT), CEA, CNRS, URA 2306, 91191
Gif-sur-Yvette, France }

\author{Hong Jiang}

\affiliation{College of Chemistry and Molecular Engineering, Peking University,
100871 Beijing, China}

\author{Silke Biermann}

\affiliation{Centre de Physique Théorique, Ecole Polytechnique, CNRS UMR 7644,
91128 Palaiseau, France}

\affiliation{Collège de France, 11 place Marcelin Berthelot, 75005 Paris, France}

\date{\today}
\begin{abstract}
Understanding the Fermi surface and low-energy excitations of iron
or cobalt pnictides is crucial for assessing electronic instabilities
such as magnetic or superconducting states. Here, we propose and implement
a new approach to compute the low-energy properties of correlated
electron materials, taking into account both screened exchange beyond
the local density approximation and local dynamical correlations.
The scheme allows us to resolve the puzzle of BaCo\textsubscript{2}As\textsubscript{2},
for which standard electronic structure techniques predict a ferromagnetic
instability not observed in nature.
\end{abstract}
\maketitle
The discovery of unconventional superconductivity in iron pnictides
and chalcogenides in 2008 has aroused strong interest into the Fermi
surfaces and low-energy excitations of transition metal pnictides
and related compounds. Angle-resolved photoemission spectroscopy (ARPES)
has been used to systematically map out quasi-particle dispersions,
and to identify electron and hole pockets potentially relevant for
low-energy instabilities \cite{ding-EPL-gap,Kaminski-BKFA-FS,Brouet-Nesting,Shin-transition,Golden-ARPES-surface,Fink-BaFe2As2,Malaeb-3d-ARPES}.
Density functional theory (DFT) calculations have complemented the
picture, yielding information about orbital characters \cite{singh-pnictides-bands},
or the dependence of the topology of the Fermi surface on structural
parameters or element substitution \cite{LaOFeAs-veronica,Mazin-Order-Parameter}.
DFT within the local density approximation (LDA) or generalized gradient
schemes has also served as a starting point for refined many-body
calculations addressing band renormalizations and quasi-particle dispersions
directly from a theoretical perspective (see e.g. Ref.{\small ~}\cite{LaOFeAs-haule-2008,cRPA-DMFT-LaOFeAs-markus,cRPA-DMFT-FeSe-markus,Valenti-LiFeAs,LaOFeAs-hansmann-2010,LaOFeAs-anisimov-2009,udyn-werner,Fang-Gutzwiller-pnictides}),
and its combination with dynamical mean field theory (LDA+DMFT) \cite{LDA+DMFT-anisimov-1997,LDA+DMFT-licht,PT-kotliar,biermann_ldadmft,held_psik,kotliar-review-DMFT,LDA-DMFT-minar}
is nowadays the state-of-the-art \textit{ab initio} many-body approach
to low-energy properties of transition metal pnictides. Despite tremendous
successes, however, limitations have also been pointed out e.g. in
the description of the Fermi surfaces. Prominent examples include
Ba(Fe,Co)\textsubscript{2}As\textsubscript{2} \cite{pnictides-QSGW-Jan,FS-Brouet} or
LiFeAs \cite{pnictides-QSGW-Jan,Valenti-LiFeAs}. Interestingly, many-body
perturbation theory approximating the self-energy by its first order
term in the screened Coulomb interaction \textit{W} (so-called ``\textit{GW}
approximation'') results in a substantially improved description:
calculations using the quasi-particle self-consistent (QS)\textit{GW}
method \cite{scGW-kotani} have pinpointed non-local self-energy corrections
to the LDA Fermi surfaces not captured in LDA+DMFT as pivotal \cite{pnictides-QSGW-Jan}.
Yet, as a perturbative method, \textit{GW} cannot be expected to describe
materials away from the weak coupling limit \cite{Thomas-PRL-GW+DMFT},
and the description of incoherent regimes \cite{cRPA-DMFT-FeSe-markus,udyn-werner}
including coherence-incoherence crossovers \cite{LaOFeAs-kotliar-2009},
local moment behavior \cite{LaOFeAs-hansmann-2010} or the subtle
effects of doping or temperature changes \cite{udyn-werner} are still
reserved for DMFT.

In this Letter, we propose and implement a new approach to the spectral
properties of correlated electron materials taking into account screened
exchange beyond the local density approximation and correlations as
described by dynamical mean field theory with frequency-dependent
local Hubbard interactions. The approach can be understood as a simplified
and extremely efficient version of the combined \textit{GW}+DMFT method
\cite{GW+DMFT-biermann}, as a non-perturbative dynamical generalization
of the popular ``COulomb-Hole-Screened-EXchange'' (COHSEX) scheme
\cite{Hedin-1965}, or as a combination of generalized Kohn-Sham schemes
\cite{Bylander-hybrid,Seidl-hybrid} with DMFT. We demonstrate the
validity of our combined ``Screened Exchange+dynamical DMFT'' (``SEx+DDMFT'')
scheme by calculating the spectral function of BaCo\textsubscript{2}As\textsubscript{2}
for which detailed ARPES results are available \cite{BaCo2As2-Nan,BaCo2As2-Dakha}.
Finally, our work sheds new light on the physical justifications of
electronic structure techniques that combine density functional with
dynamical mean field theory, by revealing a subtle error cancellation
between non-local exchange interactions and dynamical screening effects,
both neglected in standard methods.

Our target compound, BaCo$_{2}$As$_{2}$, is isostructural to the
prototypical parent compound of the so-called 122 iron-based superconductors,
BaFe$_{2}$As$_{2}$. Replacing Fe by Co, however, increases the filling
to a nominal d$^{7}$ configuration of the 3d states, with drastic
consequences: whereas compounds with filling around the d$^{6}$ configuration
exhibit characteristic power law deviations from Fermi liquid behavior
above often extremely low coherence temperatures \cite{udyn-werner},
in the d$^{7}$ compound BaCo$_{2}$As$_{2}$ ARPES identifies clearly
defined long-lived quasi-particle bands with relatively weak mass
renormalizations \cite{BaCo2As2-Nan}. Nevertheless, the electronic
structure of this compound raises puzzling questions concerning its
paramagnetic behavior. Indeed, standard electronic structure calculations
predict a huge density of states at the Fermi level, which, given
the large Stoner parameter of Co, would be expected to trigger an
instability towards a ferromagnetic state \cite{BaCo2As2-Sefat}.
The density of states of the isoelectronic compound SrCo\textsubscript{2}As\textsubscript{2}
presents the same features, but the maximum appears to be just below
the Fermi level \cite{Johnston-SrCo2As2}. Still, in SrCo\textsubscript{2}As\textsubscript{2}
-- also a paramagnet -- important antiferromagnetic fluctuations have
been measured, possibly competing with ferromagnetic order \cite{Johnston-SrCo2As2,McQueeney-SrCo2As2}.
CaCo\textsubscript{2}As\textsubscript{2}, Ca\textsubscript{0.9}Sr\textsubscript{0.1}Co\textsubscript{2}As\textsubscript{2}
and CaCo\textsubscript{1.86}As\textsubscript{2} exhibit magnetic
phases with in-plane ferromagnetism at low temperatures \cite{NLWang-CaCo2As2,Chen-CaSrCo2As2,Johnston-CaCo1.86As2}.
ARPES data of BaCo\textsubscript{2}As\textsubscript{2} show that
there is indeed a flat band (dominantly of $d_{x^{2}-y^{2}}$ character)
very close to the Fermi surface, albeit less filled than predicted
by LDA calculations \cite{BaCo2As2-Nan,BaCo2As2-Dakha}, suggesting
BaCo\textsubscript{2}As\textsubscript{2} to be on the verge of a
transition %
\footnote{Indeed, a presentation of the ARPES spectra in a second derivative
plot of the intensities reveals the presence of a flat band just above
the Fermi level \cite{BaCo2As2-Nan}, as pointed out in Ref.~\cite{BaCo2As2-Dakha}.%
}. These properties make the compound an ideal benchmark system, on
which to test new theoretical approaches.

\begin{figure}[!tph]
\begin{centering}
\includegraphics[scale=0.85]{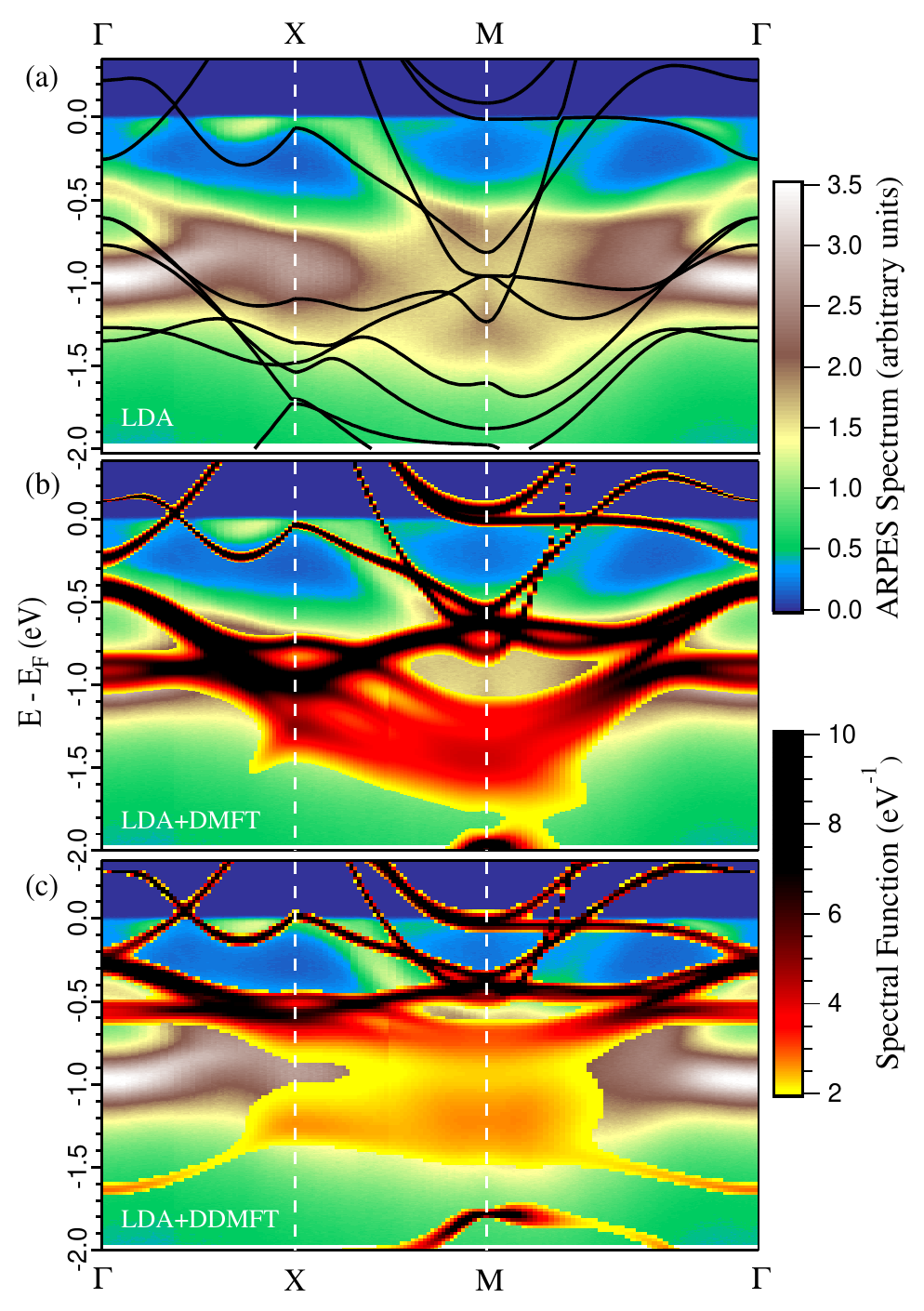} 
\par\end{centering}

\caption{{\small BaCo\textsubscript{{\small 2}}As\textsubscript{{\small 2}}
photoemission spectra, replotted from Ref.~\cite{BaCo2As2-Nan}.
Superimposed are (a) the Kohn-Sham band structure of DFT-LDA (b) spectral
function of standard LDA+DMFT {[}only those parts that exceed $2$
states/eV are shown{]} (c) spectral function within LDA+DDMFT {[}same
representation as in (b){]}. \label{LDAvsARPES}}}
\end{figure}

We start our analysis by comparing results for the spectral function
calculated within different state-of-the-art electronic structure
techniques to the ARPES spectral function of Ref.{\small ~}\cite{BaCo2As2-Nan}
(Fig.{\small ~}\ref{LDAvsARPES}). Specifically, we analyze the Kohn-Sham
band structure of DFT-LDA and the spectral functions of standard LDA+DMFT
and LDA+DMFT with frequency-dependent local Hubbard interactions $\mathcal{U}(\omega)$.
The latter scheme will be abbreviated in the following as ``LDA+DDMFT''
to stress the doubly dynamical nature of the theory, which determines
a frequency-dependent self-energy in the DMFT spirit but does so extending
DMFT to frequency-dependent interactions. Details of the scheme can
be found in Refs. \cite{udyn-michele,udyn-werner} and the supplementary
material. The effective local interactions used in the DMFT calculations
were obtained within the constrained random phase approximation in
the implementation of \cite{vaugier-crpa}. For LDA+DDMFT, the full
frequency-dependence of the monopole term, $F_{0}(\omega)$ (see supplementary)
is retained in the calculation. The effective local problem with dynamical
$\mathcal{U}$ is solved self-consistently by means of a continuous-time
Monte Carlo algorithm \cite{CTQMC-werner,CTQMC-dyn-werner} that we
have implemented within the TRIQS toolbox \cite{TRIQS-website}.

Electronic bands in the energy window between the Fermi level and
-2 eV binding energy are states of predominant Co-3\textit{d} character,
and undergo -- even in this quite moderately correlated compound --
a non-negligible band renormalization, as compared to the LDA band
structure (Fig.{\small ~}\ref{LDAvsARPES} (a)). Standard LDA+DMFT
(Fig.{\small ~}\ref{LDAvsARPES} (b)) captures this effect, leading
to a reduced bandwidth in good agreement with the ARPES results. When
dynamical screening effects are taken into account (Fig.{\small ~}\ref{LDAvsARPES}
(c)), additional renormalizations occur, corresponding to the electronic
polaron effect discussed in Ref.~\cite{udyneff-michele}, and the
overall bandwidth reduction appears to be overestimated. We will,
however, argue below that one should not conclude from this analysis
that dynamical screening effects are absent. Rather, non-local exchange
-- routinely neglected in DFT-based techniques -- reshapes and widens
the quasi-particle band structure, and the apparent success of LDA+DMFT
in obtaining the correct quasi-particle bandwidth relies on an error
cancellation when both dynamical screening and non-local exchange
are neglected in the calculation of the spectral function. We will
now substantiate this claim by explicitly including screened exchange,
and performing a DMFT calculation with fully dynamical Hubbard interactions
based on the following one-particle Hamiltonian: $H_{0}=H_{Hartree}+H_{SEx}$
where the first term denotes the Hamiltonian of the system at the
Hartree mean-field level, evaluated at the self-consistent DFT-LDA
density. $H_{SEx}$ is a screened Fock exchange term, calculated from
the Yukawa potential $\frac{e^{2}exp(-k_{TF}|r-r^{\prime}|)}{|r-r^{\prime}|}$
with screening wavevector $k_{TF}$ (see supplementary). This scheme
can be understood as the next generation after the recent LDA+DDMFT
scheme, by replacing the local Kohn-Sham exchange-correlation potential
of DFT by a non-local screened Fock exchange correction %
\footnote{We note that in SEx+DMFT non-local and dynamical renormalizations
are by construction separated on the self-energy/Hamiltonian level.
This separability was recently justified for iron pnictides \cite{pnictides-QSGW-Jan}
and found to hold also for metallic transition metal oxides \cite{SrVO3-GW+DMFT-Jan-long,SrVO3-GW+DMFT-Jan}.%
}.

\begin{figure}[!tph]
\begin{centering}
\includegraphics[scale=0.85]{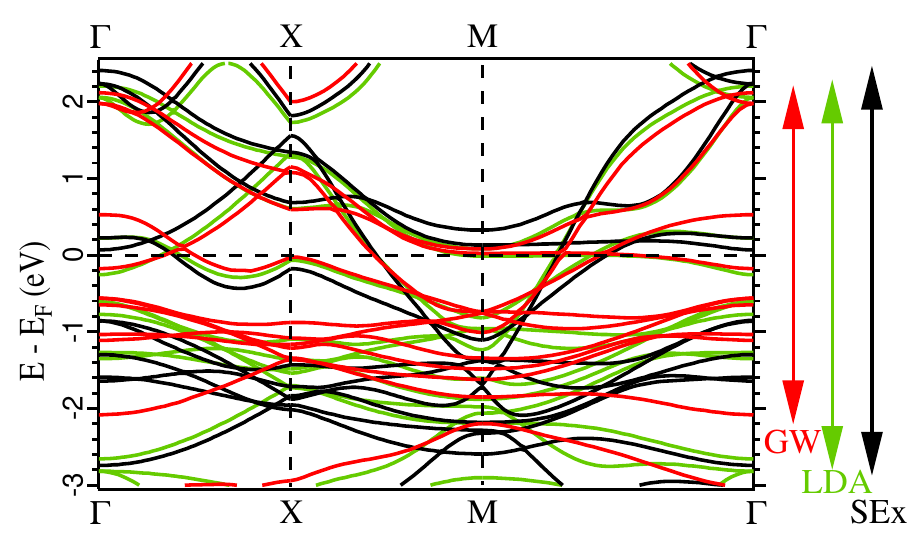} 
\par\end{centering}

\caption{{\small Comparison of bandstructures of BaCo}{\footnotesize \textsubscript{{\small 2}}}{\small As}{\footnotesize \textsubscript{{\small 2}}}{\small{}
in the $k_{z}=0$ plane calculated within QS}\textit{\small GW}{\small{}
(red), LDA (green) and Screened-Exchange (black).\label{GWvsLDAvsYukawa}}}
\end{figure}

We first analyze the band structure corresponding to $H_{0}$ alone,
in comparison to the LDA band structure and the one obtained from
QS\textit{GW} (in the implementation of \cite{SCGW-Kotani-PRB2007}),
see Fig.{\small ~}\ref{GWvsLDAvsYukawa}. As expected, the inclusion
of non-local exchange in $H_{0}$ increases the delocalization of
electrons, particularly for the bands crossing the Fermi level, and
thus widens them as compared to the LDA electronic structure. In QS\textit{GW},
this effect is overcompensated by correlation-induced band-narrowing,
and the bandwidth of Cobalt \textit{d}-like bands is about 15 $\%$
smaller than in LDA. These comparisons highlight the fact that --
taking the screened exchange band structure as a reference -- the
effective exchange-correlation potential of DFT not only incorporates
exchange (in a local fashion), but also mimics band renormalizations
due to correlations (yet without keeping track of the corresponding
spectral weight transfers). In addition, the SEx correction induces
interesting modifications of the low-energy band structure as compared
to the LDA; we will come back to this point below.

\begin{figure}[!tph]
\centering{}\includegraphics[scale=0.85]{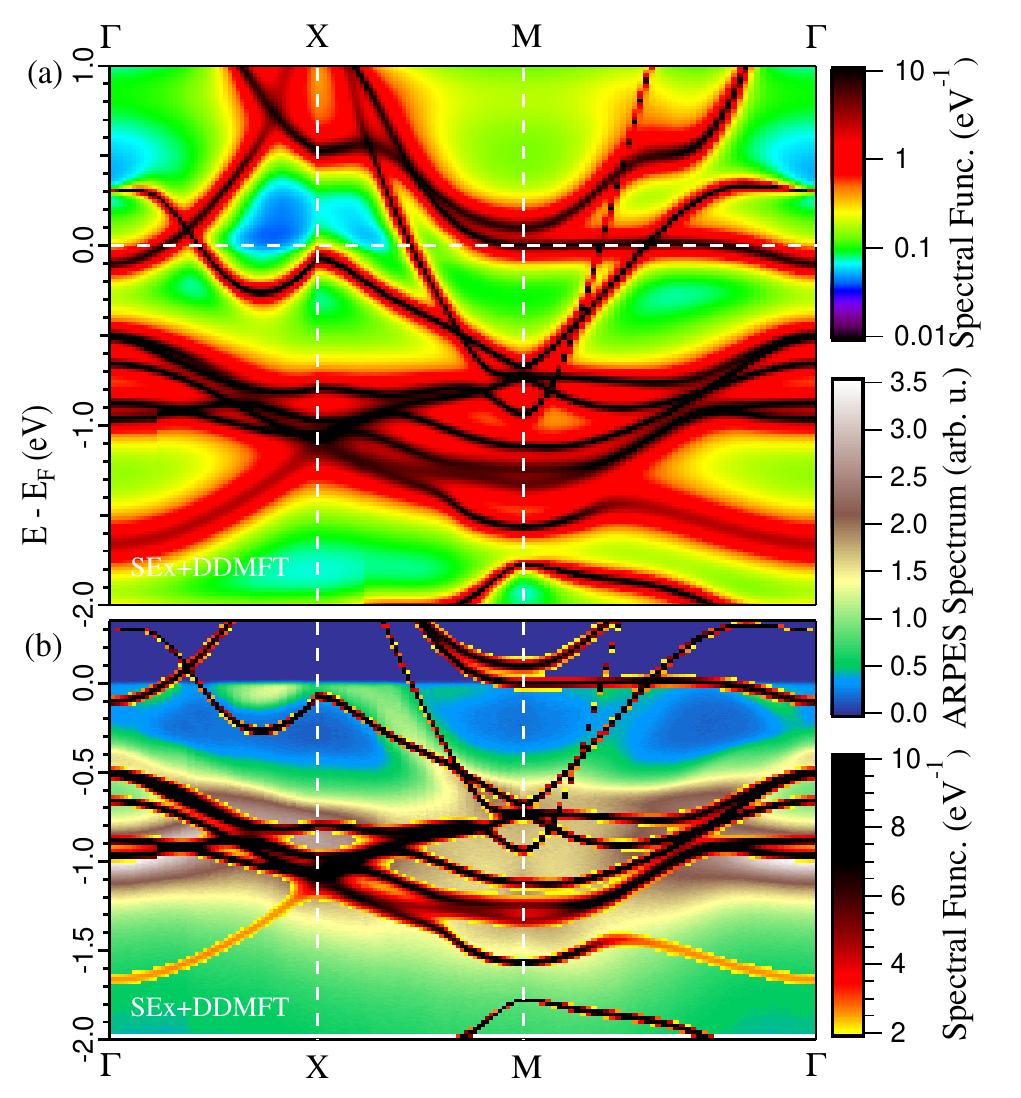} \caption{{\small BaCo}{\footnotesize \textsubscript{{\small 2}}}{\small As}{\footnotesize \textsubscript{{\small 2}}}{\small{}
within Screened Exchange+DDMFT: (a) spectral function (b) bands extracted
from the maxima of panel (a) and superimposed on ARPES data as in
Fig.~\ref{LDAvsARPES}.\label{YukvsARPES}}}
\end{figure}

\begin{figure}[!tph]
\centering{}\includegraphics[scale=0.85]{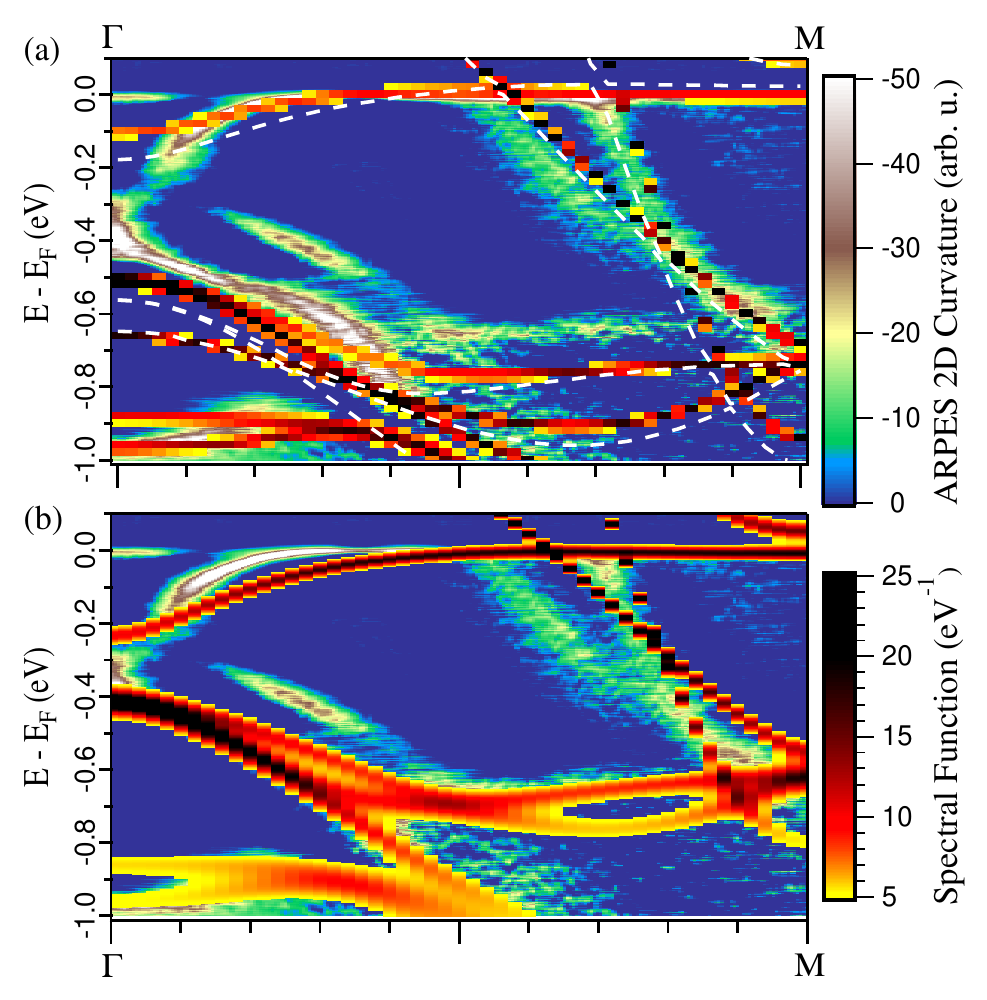} \caption{{\small Bands along the $\Gamma$M direction, extracted from the spectral
function calculated by (a) SEx+DDMFT and (b) LDA+DMFT, superimposed
on ARPES data from Ref.~\cite{BaCo2As2-Nan} (represented as a second
derivative of the photoemission intensity). The QS}\textit{\small GW}{\small{}
band structure is also given (white dashed lines).\label{GM}}}
\end{figure}

We finally turn to the results of our new scheme: Fig.{\small ~}\ref{YukvsARPES}
displays the spectral function within ``SEx+DDMFT''. Panel (b) displays
again the ARPES data, but this time the maxima of the SEx+DDMFT spectral
function are superimposed to the color spectrum. The overall spectrum
from SEx+DDMFT is very close to the experiment: bandwidth, Fermi surface
and band renormalizations close to the Fermi level are correctly predicted.

\begin{table}[!tph]
{\footnotesize \caption{{\small Number of electrons in Cobalt-}\textit{\small d}{\small{} Wannier
functions within the LDA, SEx, SEx+DDMFT, LDA+DMFT and LDA+DDMFT.
\label{electron count}}}
}{\footnotesize \par}

\centering{}{\footnotesize }%
\begin{tabular*}{1\columnwidth}{@{\extracolsep{\fill}}lccccc}
\hline 
\hline & $n_{\text{LDA}}^{e-}$ & $n_{\text{SEx}}^{e-}$ & $n_{\text{SEx+DDMFT}}^{e-}$ & $n_{\text{LDA+DMFT}}^{e-}$ & $n_{\text{LDA+DDMFT}}^{e-}$\tabularnewline
\hline 
$d_{z^{2}}$ & 1.64 & 1.66 & 1.63 & 1.61 & 1.62\tabularnewline
$d_{x^{2}-y^{2}}$ & 1.49 & 1.27 & 1.37 & 1.53 & 1.59\tabularnewline
$d_{xy}$ & 1.74 & 1.78 & 1.72 & 1.67 & 1.63\tabularnewline
$d_{xz+yz}$ & 1.69 & 1.73 & 1.68 & 1.64 & 1.62\tabularnewline
total & 8.24 & 8.16 & 8.08 & 8.09 & 8.09\\ \hline\hline%\tabularnewline
\end{tabular*}
\end{table}

The orbital-resolved electron count obtained with SEx+DDMFT is displayed
in Tab.{\small ~}\ref{electron count} and compared to the LDA, SEx,
LDA+DMFT and LDA+DDMFT electron count. The orbital polarization from
LDA is reduced by correlations, and nearly suppressed within LDA+DDMFT.
Conversely, screened exchange increases the orbital polarization,
and the final SEx+DDMFT result still displays stronger orbital polarization
than LDA. This trend can be directly related to the weakly dispersive
$d_{x^{2}-y^{2}}$ states discussed above: as in the SEx band dispersion
of Fig.{\small ~}\ref{GWvsLDAvsYukawa}, the effect of screened exchange
is to push the flat $d_{x^{2}-y^{2}}$-like band away from the Fermi
level, to the point of suppressing the electron pocket at the $\Gamma$
point. This does not correspond to the experimental spectrum, and
indeed it is corrected by including correlations. Fig.{\small ~}\ref{GM}
displays the low-energy spectra along the $\Gamma$M direction comparing
SEx+DDMFT and LDA+DMFT overlaid onto the second derivative of the
ARPES data \cite{Peng-curvature}, together with the QS\textit{GW}
band dispersion. The electron pocket at $\Gamma$ is recovered in
SEx+DDMFT, and the fraction of $d_{x^{2}-y^{2}}$ electrons increases.
Within LDA, the flat band is nearly filled along the $\Gamma$M direction,
and even more so when we take into account correlations. According
to ARPES, this flat band should be occupied only in a small electron
pocket at $\Gamma$, containing about 0.18 e\textsuperscript{-}.
This result is consistent with the absence of ferromagnetism. Indeed,
this flat band lying on the Fermi surface would imply a high density
of states at the Fermi level that could trigger a Stoner instability.
We extract from the SEx+DDMFT calculations a DOS at the Fermi energy
of 0.97 states/eV/Co/spin. Assuming a Stoner parameter of $\sim$
0.9 eV this leaves us slightly below the onset of Stoner ferromagnetism
\footnote{ However, slight electron doping would bring us close to the maximum
value of 1.04 states/eV/Co/spin that we find at $\omega=44$meV, at
the peak of the Co $d_{x^{2}-y^{2}}$ DOS, possibly triggering a ferromagnetic
instability.%
}. The QS\textit{GW} scheme also provides an overall good description,
including the position of the $d_{x^{2}-y^{2}}$ band, its filling
and its related Fermi wavevector. Taking non-local exchange into account
is thus necessary to capture the physics of BaCo$_{2}$As$_{2}$,
and our SEx+DDMFT scheme performs well for these subtle effects.

Finally, a few comments are in order to put our new computational
scheme into perspective. As far as coarse features such as the bandwidth
are concerned, standard LDA+DMFT and the new SEx+DDMFT give comparable
results, giving an \textit{a posteriori} explanation for the success
of LDA+DMFT calculations with static interactions. For total energy
calculations within DFT, it is well-known that there are subtle error
cancellations between the exchange and correlation contributions in
approximate density functionals. Here, we evidence a similar behavior
for spectral properties. The effect of dynamical screening as incorporated
in the high-energy tail of the dynamical Hubbard interaction $\mathcal{U}(\omega)$
can roughly be understood by a band narrowing by a factor $Z_{B}=\exp\left(-\int_{0}^{\infty}d\omega\Im\mathcal{U}(\omega)/(\pi\omega^{2})\right)$
\footnote{In the anti-adiabatic limit, where the characteristic frequency of
variations in $\mathcal{U}(\omega)$ is larger than the other energy
scales of the system, this statement can be made rigorous by means
of a Lang-Firsov transformation, as demonstrated in Ref.~\cite{udyneff-michele}.%
}. For BaCo$_{2}$As$_{2}$, we find dynamical screening to be non-negligible,
with $Z_{B}\sim0.6$. LDA+DDMFT doublecounts this narrowing effect,
as the bandwidth has already been decreased by correlations hidden
in the effective exchange-correlation potential of DFT, with respect
to the Hartree-Fock or SEx bandstructure (see Fig.{\small ~}\ref{GWvsLDAvsYukawa}).
Thus, starting a many-body calculation from LDA raises not only the
usual well-known double counting questions related to the energetic
position of correlated versus itinerant states, but even more serious
ones related to the double counting of screening processes. Our SEx+DDMFT
scheme avoids these issues, providing a more solid foundation for
the investigation of dynamical screening effects. On a more pragmatic
level, the similarity of the LDA+DMFT and SEx+DDMFT spectral functions
also suggests that error cancellations between dynamical screening
and non-local exchange, both absent in LDA+DMFT, make this scheme
suitable at least for questions concerning the overall bandwidth reduction
of correlated electron systems. Finer details related to the very
low energy behavior or Fermi surface topologies, on the other hand,
might require explicit exchange corrections as introduced in the present
work.

In summary, we have shown that screened exchange combined with dynamical
correlations provides an excellent description of the low-energy physics
in BaCo\textsubscript{2}As\textsubscript{2}. In contrast to perturbative
schemes, it can be expected that our non-perturbative method can be
extended to regimes with arbitrarily strong correlations, making it
a promising tool for probing the finite temperature normal state of
iron pnictide and chalcogenide superconductors. For BaCo$_{2}$As$_{2}$,
we show that the flat $d_{x^{2}-y^{2}}$-band in the immediate vicinity
of the Fermi level is extremely sensitive to an accurate treatment
of screened exchange, and that this effect is key to the paramagnetic
nature of the compound. Pump-probe photoemission would be useful to
experimentally locate the flat band and guide the search for new ways
to tune its exact energetic position, thus directly playing on possible
Fermi surface instabilities.

We acknowledge useful discussions with V. Brouet, T. Miyake and the
authors of Ref.{\small ~}\cite{BaCo2As2-Nan}. This work was supported
by the French ANR under project PNICTIDES, IDRIS/GENCI under projects
91393 and 96493, the Cai Yuanpei program, and the European Research
Council (projects number
617196 and 278472). JMT acknowledges hospitality of CPHT within a
CNRS visiting position. HJ acknowledges the support by the National
Natural Science Foundation of China (Projects Nos. 20973009 and 21173005).

\bibliographystyle{plain}

%%%%%%%%%% Merge with supplemental materials %%%%%%%%%%
\widetext
\clearpage
\begin{center}
\textbf{\large Dynamical correlations and screened exchange on the experimental
bench: spectral properties of the cobalt pnictide BaCo\textsubscript{2}As\textsubscript{2}
- Supplementary material}
\end{center}
\twocolumngrid
%%%%%%%%%% Merge with supplemental materials %%%%%%%%%%
%%%%%%%%%% Prefix a "S" to all equations, figures, tables and reset the counter %%%%%%%%%%
\setcounter{equation}{0}
\setcounter{figure}{0}
\setcounter{table}{0}
\setcounter{page}{1}
\makeatletter
\renewcommand{\theequation}{S\arabic{equation}}
\renewcommand{\thefigure}{S\arabic{figure}}
\renewcommand{\bibnumfmt}[1]{[S#1]}
\renewcommand{\citenumfont}[1]{S#1}
%%%%%%%%%% Prefix a "S" to all equations, figures, tables and reset the counter %%%%%%%%%%

In this supplementary material, we give further details explaining
the motivation, implementation and technical aspects of the proposed
combined Screened Exchange Dynamical Mean Field scheme with Dynamical
interactions ``SEx+DDMFT''.

\section{The combined Screened Exchange Dynamical Mean Field (SEx+DMFT) scheme}

\subsection{Motivation and general idea}

The screened exchange dynamical mean field technique ``SEx+DDMFT''
is motivated by the observation that the first non-local corrections
to standard LDA+DMFT can be understood as screened exchange terms.
Indeed, within the \textit{GW} approximation (GWA) it has been shown
recently \cite{pnictides-QSGW-Jan-s,SrVO3-GW+DMFT-Jan-long-s} that for
whole classes of materials (the focus was in particular on iron pnictides),
the \textit{GW} correction to LDA can be split into two contributions:
a dynamical but local self-energy $\Sigma_{loc}(\omega)$ and a k-dependent
but static self-energy $\Sigma_{nloc}(k)$. Obviously, if such a separation
were strictly valid over all energies, the static contribution would
be given by the Hartree-Fock self-energy, since the dynamical part
of the self-energy vanishes at high frequency. However, in reality,
such a decomposition holds to a good approximation in the low-energy
regime that we are interested in, with a static offset $\Sigma_{nloc}(k)$
which is quite different from the Fock exchange term. Here, we identify
this contribution as a screened exchange self-energy, leading to a
decomposition of the \textit{GW} self-energy $\Sigma_{GW}(k,\omega)=i\sum_{\nu,q}G(k+q,\omega+\nu)W(q,\nu)$
into $\Sigma_{GW}(k,\omega)=GW(\nu=0)+[GW]_{local}$
where the first term is a screened exchange contribution arising from
the screened interaction $W$ evaluated at zero frequency. In the
practical implementation we replace it by its long-wavelength limit:
in this $q\rightarrow0$ limit, the RPA-screened Coulomb potential
$W$ reduces to a simple Yukawa-form, which we discuss below. Within
the GWA, the dynamical local term is simply given by a \textit{GW}
self-energy evaluated using a local propagator $G_{local}$ and the
local screened Coulomb interaction $W_{local}$. In the SEx+DDMFT
scheme, however, we go a decisive step further and evaluate this term
from a dynamical impurity problem as in DDMFT. In contrast to the
GWA, it is therefore determined non-perturbatively. For weakly correlated
materials such as BaCo$_{2}$As$_{2}$ we expect the GWA to be quite
accurate, and the non-perturbative and the perturbative evaluation
should qualitatively coincide. This corresponds indeed to our finding
of SEx+DDMFT results being close to the \textit{GW} ones (see main
text). The SEx+DDMFT is however not limited to this case, making it
a promising perspective for investigations in the strongly correlated
metal regime.

\subsection{Formulation}

Thanks to its static nature, the screened exchange self-energy correction
to the LDA Hamiltonian can be directly incorporated into a revised
one-particle Hamiltonian $H_{0}=H_{LDA}-V_{xc}^{LDA}+V_{SEx}$. The
SEx+DDMFT scheme then consists in performing DDMFT starting from the
general multi-orbital Hamiltonian $H=H_{0}+H_{int}-H_{dc}$ where
$H_{int}$ is the dynamical generalization of the usual Hubbard and
Hund interaction terms. To indeed be able to write it in a Hamiltonian
fashion, we decompose it into a static part which we restrict to density-density-interactions
$H_{intstat}=\frac{1}{2}\sum_{i}\sum_{m\sigma\neq m'\sigma'}U_{m\sigma m'\sigma'}(\nu=\infty)n_{im\sigma}n_{im'\sigma'}$
(where the sums $i,j$ run over the Co-3\textit{d} orbitals) and an
additional part introducing screening bosons and their coupling to
the physical electrons: $H_{screening}=\sum_{i}\int_{\omega}d\omega\left[\lambda_{\omega}\sum_{m\sigma}n_{im\sigma}(b_{i,\omega}^{\dagger}+b_{i,\omega})+\omega\left(b_{i,\omega}^{\dagger}b_{i,\omega}+\frac{1}{2}\right)\right]$.
Such a formulation has been discussed in the recent generalization
of LDA+DMFT to dynamical Hubbard interactions in Refs.\cite{CTQMC-dyn-werner-s,udyneff-michele-s}.
Alternatively, if an action formulation is chosen, $H_{int}$ corresponds
to the Hubbard interactions where the monopole term (the Slater integral
$F_{0}$) is frequency-dependent. $H_{dc}$ is the double-counting
correction introduced in order to remove from the screened exchange
Hamiltonian those contributions of the interaction that are explicitly
treated in a many-body fashion in the form of $H_{int}$. For details,
see Section \ref{dc}.

\subsection{Screened exchange term and screening length}

The screened exchange contribution $H_{SEx}$ is calculated as Fock
exchange with the screened potential $W^{RPA}(q\rightarrow0,\nu=0)\rightarrow\frac{e^{2}}{4\pi\epsilon_{0}}\frac{4\pi}{q^{2}+k_{TF}^{2}}$.
Here, $k_{TF}$ is the Thomas-Fermi wavevector or inverse screening
length. It can be expressed as $k_{TF}^{2}\equiv-e^{2}\epsilon_{0}^{-1}P^{0}(q=0,\nu=0)$
where -- in a metallic system such as BaCo$_{2}$As$_{2}$ -- the
static polarization $P^{0}(0,0)$ is well approximated by the one
of the homogeneous electron gas at low temperature $k_{B}T\ll\epsilon_{F}$:
$P^{0}(0,0)\rightarrow-\rho(\epsilon_{F})$. The problem of calculating
the screened exchange term therefore boils down to determining the
effective screening length. Starting from the LDA DOS (6.94 states/eV/unit
cell), we obtain a screening wavevector of 1.88 $a_{0}^{-1}$, with
$a_{0}$ the Bohr radius. However, as argued above, one of the main
consequences of the electronic exchange and correlation effects is
precisely the rearrangement of the electronic states in the immediate
proximity of the Fermi level, and a concomitant reduction of the density
of states. Indeed, a SEx+DDMFT calculation with the LDA screening
length results in a reduction of states at the Fermi level, given
by a zero-frequency spectral function of 4.66 states/eV/unit cell,
corresponding to a screening wavevector of 1.54 $a_{0}^{-1}$. It
thus becomes clear that the true challenge consists in a \textit{self-consistent}
determination of the effective screening length. For computational
reasons, however, and since the precise value of the DOS at the Fermi
level is -- for the same reasons as the subtle Fermi surface modifications
discussed above -- a quantity very hard to converge, we mimic convergence
by choosing an ad hoc screening wavevector of 1.33 $a_{0}^{-1}$,
corresponding to half the LDA-DOS. Since this is approximately the
value we eventually find for the value of the spectral function on
the Fermi level (4.26 states/eV/cell -- which corresponds to about
1.47 $a_{0}^{-1}$), this choice can be considered as a poor man's
self-consistency.

For the screened Hartree-Fock calculation performed within Wien2k
\cite{blaha_wien2k-s,wien-hf-Tran-s}, we calculate the matrix-elements
for a second variational procedure on 150 bands, on a grid of 7x7x7
k-points.

\subsection{Interactions}

\subsubsection{Static Hubbard and Hund Interactions}

The interactions are determined from the constrained random phase
approximation (cRPA) in the implementation of \cite{vaugier-crpa-s}.
In the zero-frequency limit, we obtain the Slater integrals $F^{0}(0)=2.9$,
$F^{2}(0)=6.9$ and $F^{4}(0)=5.1$ calculated on a grid of 4x4x4
k-points, which we use to determine the orbital-dependent coefficients
$U_{mm'}$ of density-density interactions, including Hund's exchange
$J_{mm'}$, in a standard Slater parametrisation \cite{loig-tesis-s}.
We find $J=0.86$ and the interaction matrices in the cubic basis
$d_{z^{2}}$, $d_{x^{2}-y^{2}}$, $d_{xy}$, $d_{xz}$, $d_{yz}$
at $\omega=0$: 
\[
\bar{U}_{mm'}^{\sigma\sigma}|_{\textrm{Slater}}=\left(\begin{array}{ccccc}
0 & 1.66 & 1.66 & 2.41 & 2.41\\
1.66 & 0 & 2.66 & 1.91 & 1.91\\
1.66 & 2.66 & 0 & 1.91 & 1.91\\
2.41 & 1.91 & 1.91 & 0 & 1.91\\
2.41 & 1.91 & 1.91 & 1.91 & 0
\end{array}\right)
\]
 
\[
\bar{U}_{mm'}^{\sigma\bar{\sigma}}|_{\textrm{Slater}}=\left(\begin{array}{ccccc}
3.87 & 2.40 & 2.40 & 2.90 & 2.90\\
2.40 & 3.87 & 3.06 & 2.56 & 2.56\\
2.40 & 3.06 & 3.87 & 2.73 & 2.56\\
2.90 & 2.56 & 2.56 & 3.87 & 2.56\\
2.90 & 2.56 & 2.56 & 2.56 & 3.87
\end{array}\right).
\]

\subsubsection{Dynamical interactions}

The cRPA method also gives us access to the frequency-dependence of
the interactions. 
\begin{figure}[!tph]
\begin{centering}
\includegraphics{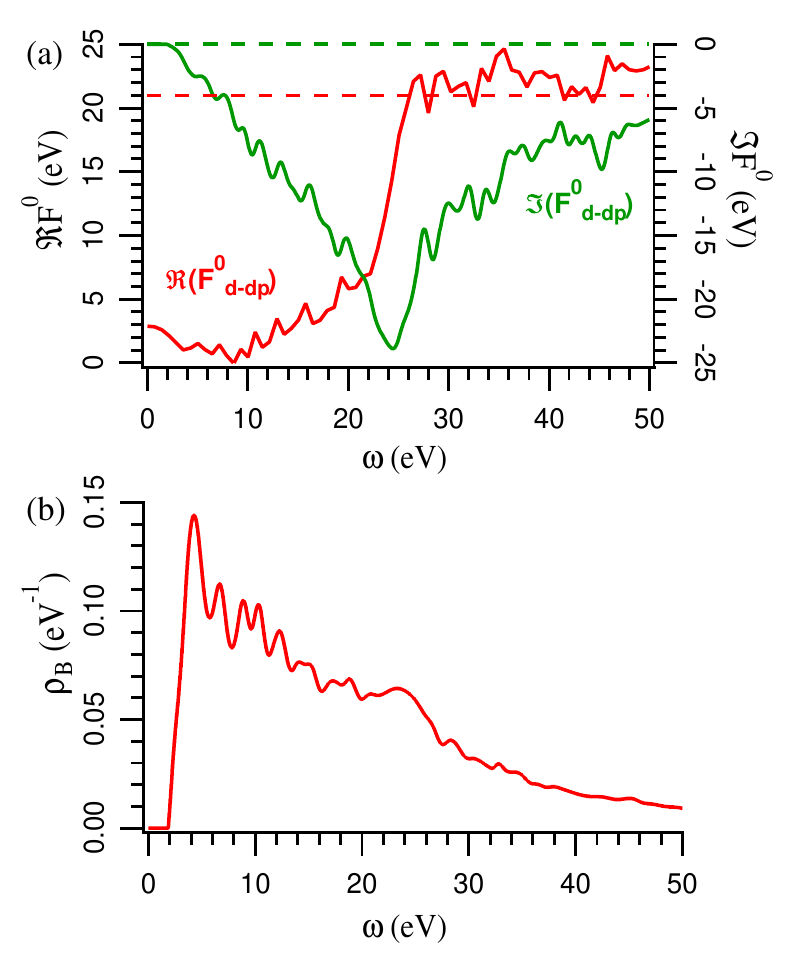}
\par\end{centering}

\caption{(a)Frequency-dependent real and imaginary parts of the monopole screened
Coulomb interaction $F_{d-dp}^{0}(\omega)$; (b)Corresponding bosonic
spectral function $\rho_{B}(\omega)$.\label{U}}
\end{figure}

The full frequency-dependent monopole integral $F^{0}(\omega)$ is
displayed in Figure \ref{U}, along with the corresponding bosonic
spectral function $\rho_{B}(\omega)$ with the notations of Ref. \cite{udyn-werner-s}.
The imaginary part of $F^{0}$ has a minimum around 22 eV, which corresponds
to the main plasmon. At this frequency the real part of $F^{0}$ is
multiplied by 4, going from a low-frequency regime to a high-frequency
regime. Still, a non-trivial frequency dependence persists at lower
frequencies, and that is why the spectrum of $\rho_{B}$ shows many
features and a broad repartition of the spectral weight. Formally,
the bosonic renormalization factor (within the notations of Ref. \cite{udyneff-michele-s})
would be $Z_{B}$=0.59.

\subsection{Double-counting}

\label{dc}

The Hubbard Hamiltonian used in DMFT contains a Hartree term. However,
the Hartree energy from the input (whether it is LDA or calculated
from a Yukawa potential) is already taken into account in the DFT
part. For this reason, one needs to introduce a correction term introduced
to avoid double-counting part of the Coulomb interactions. Here, we
use an orbital-dependent double counting term derived as a mean field
approximation of the Hubbard terms: $V_{DC}(m,\sigma)=\sum_{m',\sigma'\neq m,\sigma}U(m\sigma,m'\sigma')<n_{m'\sigma'}>_{DFT}$.
We note that in the absence of orbital polarization one would recover
the usual ``around mean field'' (AMF) formula \cite{cLDA-anisimov-1991-s}.
The values of $<n_{m'\sigma'}>_{DFT}$ are taken from the DFT starting
point -- the LDA or SEx calculation.

\begin{table}[!tph]
{\footnotesize \caption{{\footnotesize Orbital-resolved value of double-counting for LDA and
SEx, and effective mass of Cobalt-}\textit{\footnotesize d}{\footnotesize{}
Wannier functions in SEx+DDMFT, LDA+DMFT and LDA+DDMFT.\label{DC and m*}}}
}{\footnotesize \par}

\centering{}{\footnotesize }%
\begin{tabular*}{1\columnwidth}{@{\extracolsep{\fill}}cccccc}
 &  &  &  &  & \tabularnewline
\hline 
\hline 
 & $DC_{LDA}$ & $DC_{SEx}$ & $m_{SEx+U(\omega)}^{*}$ & $m_{LDA+U(0)}^{*}$ & $m_{LDA+U(\omega)}^{*}$\tabularnewline
\hline 
$d_{z^{2}}$ & 18.665 & 18.552 & 1.74 & 1.26 & 1.95\tabularnewline
$d_{x^{2}-y^{2}}$ & 18.493 & 18.159 & 1.86 & 1.27 & 1.90\tabularnewline
$d_{xy}$ & 18.719 & 18.628 & 1.68 & 1.25 & 1.94\tabularnewline
$d_{xz}/d_{yz}$ & 18.599 & 18.415 & 1.68 & 1.24 & 1.90\tabularnewline
\hline 
\hline 
 &  &  &  &  & \tabularnewline
\end{tabular*}
\end{table}

The values of the double counting term are displayed in Table \ref{DC and m*}.

\subsection{Screened EXchange+Dynamical Mean Field Theory: Link to \textit{GW}+DMFT
and COHSEX}

The above decomposition of the self-energy, which essentially consists
in singling out the local term, is reminiscent of the combined \textit{GW}+DMFT
scheme\cite{GW+DMFT-biermann-s}, in which local and non-local parts
of the self-energy are treated within DMFT and \textit{GW} respectively.
Indeed, the SEx+DMFT scheme can be understood as a simplified version
of \textit{GW}+DMFT, where the non-local self-energy is approximated
by the screened exchange term.

Alternatively, the SEx+DDMFT scheme can also be understood as a dynamical
generalization of Hedin's Coulomb-hole-screened-exchange (COHSEX)
approximation \cite{Hedin-review-s}. In that scheme, a static approximation
to the self-energy is constructed from two terms: first, a screened
exchange term analogous to the one we treat here, and second the Coulomb-hole,
a local correction expressing the suppression of the charge-charge
correlator in the vicinity of an electron. In SEx+DDMFT, the latter
term is determined in a dynamical and non-perturbative manner from
a local impurity model (that is, by DMFT with dynamical interactions).

\section{Additional results on BaCo\textsubscript{2}As\textsubscript{2}}

In this section, we give additional information on our SEx+DDMFT results,
concerning the momentum-integrated spectral function as well as the
very low-energy behavior related to the flat band at the Fermi surface.

\subsection{Momentum-integrated cobalt-3d spectral function}

\begin{figure}[!tph]
\begin{centering}
\includegraphics[scale=0.75]{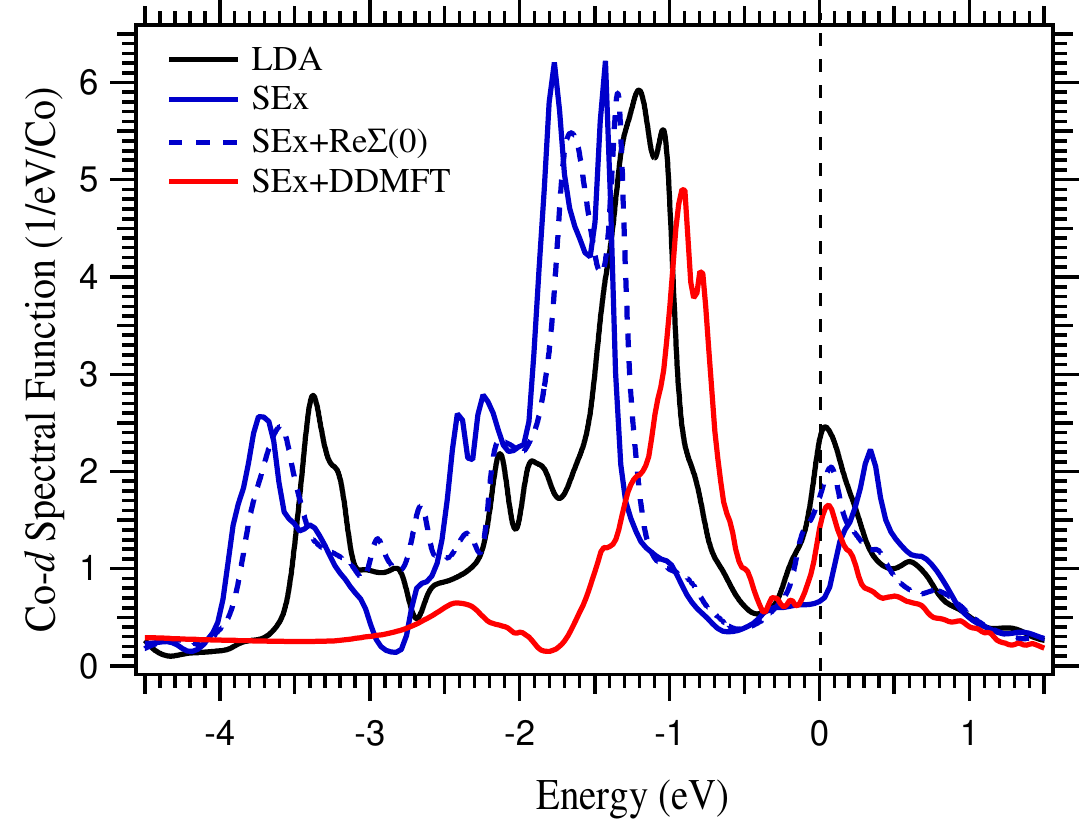} 
\par\end{centering}

\caption{Cobalt-\textit{d} spectral function from SEx+DDMFT (red) compared
to the Co-\textit{d} density of states within LDA (black), SEx (solid
blue lines), and SEx shifted by the real part of the self-energy at
zero frequency (dashed blue lines).\label{DOS}}
\end{figure}

Figure \ref{DOS} displays the momentum-integrated Co-3\textit{d}
spectral function within the different schemes: LDA, SEx, and SEx+DDMFT.
Also shown is the density of states corresponding to an intermediate
level of sophistication, namely resulting from SEx shifted by the
orbital-dependent self-energy matrix at zero frequency. Although a
purely auxiliary quantity, this density of states is interesting,
since the full SEx+DDMFT spectral function differs from it only through
a self-energy correction that vanishes on the Fermi level. For this
reason, the value on the Fermi level of the two quantities coincide,
up to finite temperature effects encoded in the imaginary part of
the self-energy (and, of course, numerical limitations). However,
this value is strongly reduced compared to the initial LDA DOS, as
a consequence of the exchange-enlarged bandwidth. The new value is
consistent with the absence of ferromagnetism, as discussed in the
main text.

\subsection{Low-energy behavior along the $\Gamma$M direction}

\begin{figure}[!tph]
\begin{centering}
\includegraphics[scale=0.8]{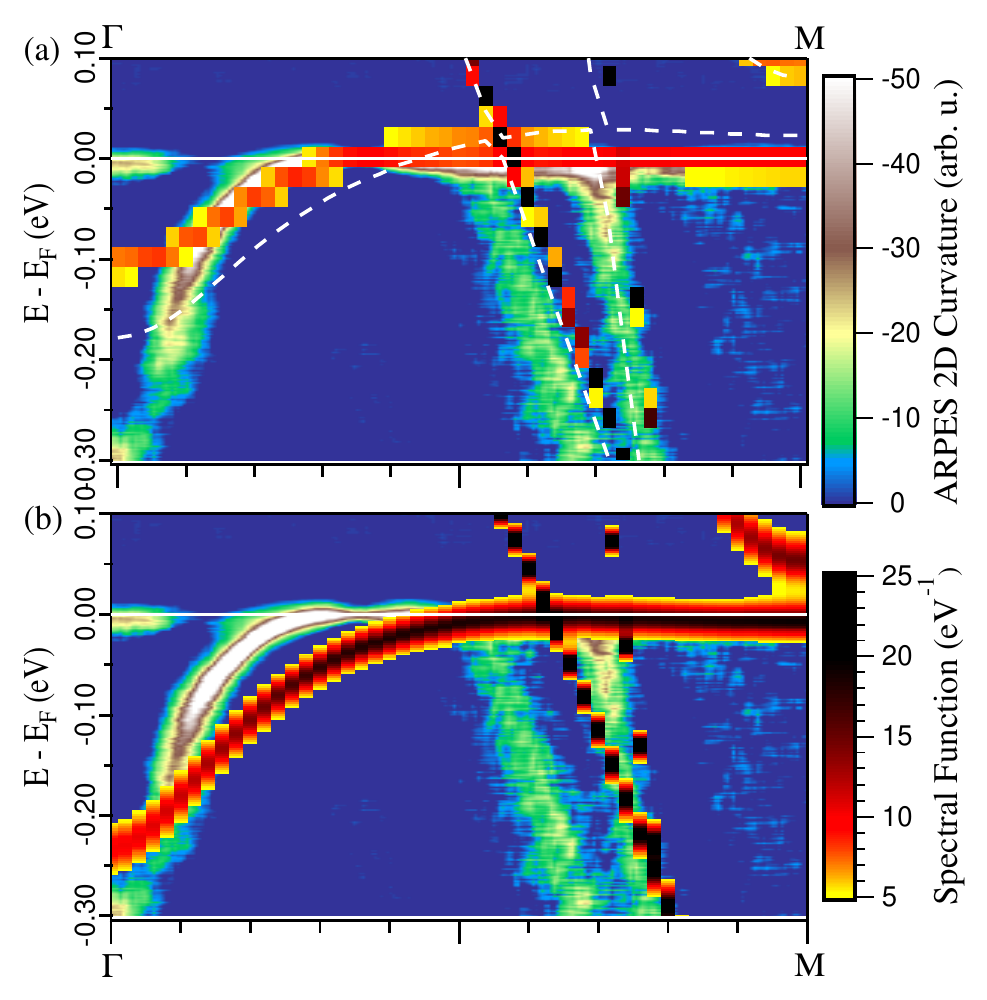} 
\par\end{centering}

\caption{Quasi-particle dispersion near the Fermi surface along the $\Gamma$M
direction, extracted from the spectral function calculated by (a)
SEx+DDMFT and (b) LDA+DMFT, superimposed on ARPES data as in Fig.~4
of the main text. The QS\textit{GW} band structure is also given (white
dashed lines). The Fermi level is indicated by the solid white line.\label{GM_zoom}}
\end{figure}

On Fig. \ref{GM_zoom}, we provide a zoom of the SEx+DDMFT and LDA+DMFT
spectral functions in the very low-energy region around the Fermi
level. Even though at this scale the numerical precision becomes challenging,
the momentum-dependent corrections to the flat low-energy band and
the Fermi surface induced by the screened exchange terms are clearly
visible.

\subsection{Co-$d_{x^{2}-y^{2}}$ orbital character}

\vspace{-0.5cm}

\begin{figure}[!tph]
\begin{centering}
\includegraphics[scale=0.8]{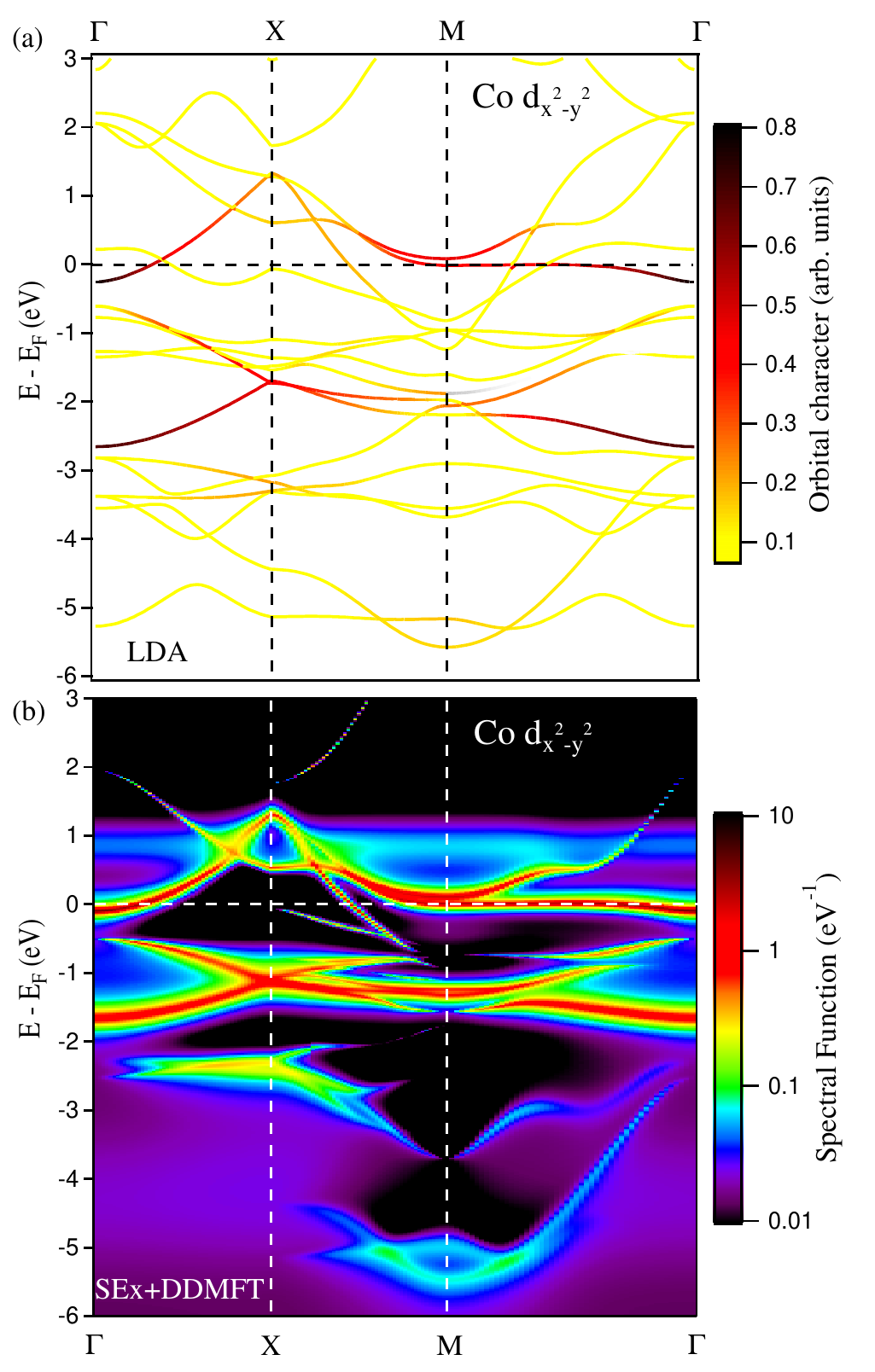} 
\par\end{centering}

\caption{Co-$d_{x^{2}-y^{2}}$ orbital character in (a) LDA bands and (b) SEx+DDMFT
spectral function.\label{dx2-y2}}
\end{figure}

As a further support for the interesting physics related to the Co-$d_{x^{2}-y^{2}}$
orbital, we display on Figure \ref{dx2-y2} the orbital-resolved LDA
band structure and SEx+DDMFT spectral function for this orbital character.
We can see that indeed the lowest Co-$d$ band is in majority of $d_{x^{2}-y^{2}}$
character, as well as the flat band near the Fermi level. This is
in contrast to the iron pnictides, in which the Fermi level lies in-between
those bonding and non-bonding $d_{x^{2}-y^{2}}$ bands, such that
there are no $d_{x^{2}-y^{2}}$ states at the Fermi surface. On the
other hand, the electron pockets near the M point are, similarly to
iron pnictides, of $d_{xy}$ and $d_{xz+yz}$ character. As a result,
the extension of the electron pockets near M and the contraction of
the electron pocket near $\Gamma$ caused by the non-local screened
exchange is at the origin of the reduction of the number of electrons
in the $d_{x^{2}-y^{2}}$ orbital.

\subsection{Further perspectives}

We finally comment on two remaining discrepancies between our theoretical
spectral functions and the ARPES spectra of Fig.~4 (main text). At
one fourth of the $\Gamma$M path, the isolated feature at -0.4 eV
is probably due to spin-orbit coupling, as a QS\textit{GW} calculation
with spin-orbit interactions (not shown) lets us foresee. Around the
X point (see Fig.~3(b) of the main text), our theoretical data does
not match the Fermi surface deduced from the ARPES spectrum. However,
we note that this point is not a high-symmetry point and it is consequently
difficult to identify it with certainty. Moreover, the ARPES cuts
correspond to a single photon energy, and in the free electron final
state approximation for a given photon and binding energy the sum
$k_{\perp}^{2}+k_{\parallel}^{2}$ is constant. This means in our
case that for the ``$k_{z}=0$'' cut the $k_{z}$ at the M point
is actually lower by about 0.32$\pi/c'$ compared to the $\Gamma$
point and at the X point it is about 0.16$\pi/c'$ lower (where $c'$
is the distance between two CoAs planes). Given the precision of our
comparison, we consider that we might be within the error bars. Further
combined ARPES and SEx+DDMFT studies will be needed to assess the
respective limitations in precision inherent to theory and experiments.

\section{Application to SrVO\textsubscript{3}}

\begin{figure}[!tph]
\begin{centering}
\includegraphics[scale=0.85]{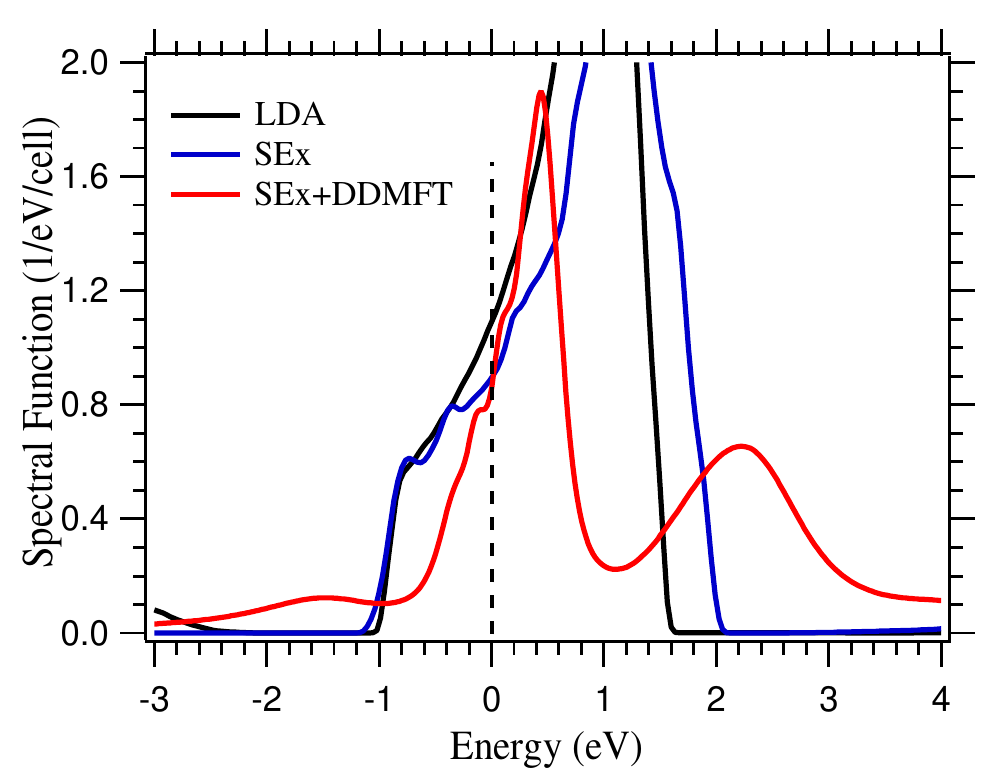}
\par\end{centering}

\caption{Comparison of the SEx+DDMFT spectral function of the $t_{2g}$ states
of SrVO\textsubscript{3} (red) with the $t_{2g}$ density of states
within LDA (black) and SEx (blue).\label{SrVO3}}
\end{figure}

As support to our new SEx+DDMFT scheme, we provide in this paragraph
a benchmark on a classical transition metal oxide: SrVO\textsubscript{3},
a cubic perovskite compound in the correlated metal regime, can be
considered as a drosophila system of strongly correlated materials.
We use the dynamical interactions as calculated within cRPA (see e.g.
\cite{SrVO3-GW+DMFT-Jan-long-s}), corresponding to a zero-frequency
value of $U(\omega=0)=3.5$ eV, and a dynamical tail leading to a
bosonic renormalization factor of $Z_{B}=0.7$. The screening wavevector
is estimated to be $k_{TF}=1.1$ $a_{0}^{-1}$ from the SEx density of
states. For SrVO\textsubscript{3}, with three fully degenerate $t_{2g}$-states,
this value corresponds by construction to the self-consistent density
of states (differing only slightly from the LDA screening wavevector of
$k_{TF}=1.2$ $a_{0}^{-1}$). In Fig. \ref{SrVO3}, we display the density
of states of the $t_{2g}$ orbitals within DFT-LDA and SEx compared
to the spectral function obtained within SEx+DDMFT. We observe the
well-known three-peak structure with a renormalized quasi-particle
peak and upper and lower Hubbard bands. However, at variance with
LDA+DMFT spectra, the position of the upper Hubbard band is decreased
in energy, locating it at energies slightly larger than 2 eV. This
is consistent with the findings in \textit{GW}+DMFT calculations \cite{SrVO3-GW+DMFT-Jan-long-s},
which predict an upper Hubbard band at this energy, identifying the
pronounced peak at 2.7 eV seen in inverse photoemission spectroscopy
\cite{SrVO3-morikawa-s} as resulting from V-$e_{g}$ states. The compensation
effect between non-local screened-exchange and the dynamical tail
discussed above for BaCo\textsubscript{2}As\textsubscript{2} is
confirmed also in this case. A more detailed account of these calculations will be published
in Ref. \cite{SrVO3-Ambroise-s}.

\bibliographystyle{plain}

\begin{thebibliography}{52}
\expandafter\ifx\csname natexlab\endcsname\relax\def\natexlab#1{#1}\fi
\expandafter\ifx\csname bibnamefont\endcsname\relax
  \def\bibnamefont#1{#1}\fi
\expandafter\ifx\csname bibfnamefont\endcsname\relax
  \def\bibfnamefont#1{#1}\fi
\expandafter\ifx\csname citenamefont\endcsname\relax
  \def\citenamefont#1{#1}\fi
\expandafter\ifx\csname url\endcsname\relax
  \def\url#1{\texttt{#1}}\fi
\expandafter\ifx\csname urlprefix\endcsname\relax\def\urlprefix{URL }\fi
\providecommand{\bibinfo}[2]{#2}
\providecommand{\eprint}[2][]{\url{#2}}

\bibitem[{\citenamefont{Ding et~al.}(2008)\citenamefont{Ding, Richard,
  Nakayama, Sugawara, Arakane, Sekiba, Takayama, Souma, Sato, Takahashi
  et~al.}}]{ding-EPL-gap}
\bibinfo{author}{\bibfnamefont{H.}~\bibnamefont{Ding}},
  \bibinfo{author}{\bibfnamefont{P.}~\bibnamefont{Richard}},
  \bibinfo{author}{\bibfnamefont{K.}~\bibnamefont{Nakayama}},
  \bibinfo{author}{\bibfnamefont{K.}~\bibnamefont{Sugawara}},
  \bibinfo{author}{\bibfnamefont{T.}~\bibnamefont{Arakane}},
  \bibinfo{author}{\bibfnamefont{Y.}~\bibnamefont{Sekiba}},
  \bibinfo{author}{\bibfnamefont{A.}~\bibnamefont{Takayama}},
  \bibinfo{author}{\bibfnamefont{S.}~\bibnamefont{Souma}},
  \bibinfo{author}{\bibfnamefont{T.}~\bibnamefont{Sato}},
  \bibinfo{author}{\bibfnamefont{T.}~\bibnamefont{Takahashi}},
  \bibnamefont{et~al.}, \bibinfo{journal}{EPL} \textbf{\bibinfo{volume}{83}},
  \bibinfo{pages}{47001} (\bibinfo{year}{2008}).

\bibitem[{\citenamefont{Liu et~al.}(2008)\citenamefont{Liu, Samolyuk, Lee, Ni,
  Kondo, Santander-Syro, Bud'ko, McChesney, Rotenberg, Valla
  et~al.}}]{Kaminski-BKFA-FS}
\bibinfo{author}{\bibfnamefont{C.}~\bibnamefont{Liu}},
  \bibinfo{author}{\bibfnamefont{G.~D.} \bibnamefont{Samolyuk}},
  \bibinfo{author}{\bibfnamefont{Y.}~\bibnamefont{Lee}},
  \bibinfo{author}{\bibfnamefont{N.}~\bibnamefont{Ni}},
  \bibinfo{author}{\bibfnamefont{T.}~\bibnamefont{Kondo}},
  \bibinfo{author}{\bibfnamefont{A.~F.} \bibnamefont{Santander-Syro}},
  \bibinfo{author}{\bibfnamefont{S.~L.} \bibnamefont{Bud'ko}},
  \bibinfo{author}{\bibfnamefont{J.~L.} \bibnamefont{McChesney}},
  \bibinfo{author}{\bibfnamefont{E.}~\bibnamefont{Rotenberg}},
  \bibinfo{author}{\bibfnamefont{T.}~\bibnamefont{Valla}},
  \bibnamefont{et~al.}, \bibinfo{journal}{Phys. Rev. Lett.}
  \textbf{\bibinfo{volume}{101}}, \bibinfo{pages}{177005}
  (\bibinfo{year}{2008}).

\bibitem[{\citenamefont{Brouet et~al.}(2009)\citenamefont{Brouet, Marsi,
  Mansart, Nicolaou, Taleb-Ibrahimi, Le~F\`evre, Bertran, Rullier-Albenque,
  Forget, and Colson}}]{Brouet-Nesting}
\bibinfo{author}{\bibfnamefont{V.}~\bibnamefont{Brouet}},
  \bibinfo{author}{\bibfnamefont{M.}~\bibnamefont{Marsi}},
  \bibinfo{author}{\bibfnamefont{B.}~\bibnamefont{Mansart}},
  \bibinfo{author}{\bibfnamefont{A.}~\bibnamefont{Nicolaou}},
  \bibinfo{author}{\bibfnamefont{A.}~\bibnamefont{Taleb-Ibrahimi}},
  \bibinfo{author}{\bibfnamefont{P.}~\bibnamefont{Le~F\`evre}},
  \bibinfo{author}{\bibfnamefont{F.}~\bibnamefont{Bertran}},
  \bibinfo{author}{\bibfnamefont{F.}~\bibnamefont{Rullier-Albenque}},
  \bibinfo{author}{\bibfnamefont{A.}~\bibnamefont{Forget}}, \bibnamefont{and}
  \bibinfo{author}{\bibfnamefont{D.}~\bibnamefont{Colson}},
  \bibinfo{journal}{Phys. Rev. B} \textbf{\bibinfo{volume}{80}},
  \bibinfo{pages}{165115} (\bibinfo{year}{2009}).

\bibitem[{\citenamefont{Shimojima et~al.}(2010)\citenamefont{Shimojima,
  Ishizaka, Ishida, Katayama, Ohgushi, Kiss, Okawa, Togashi, Wang, Chen
  et~al.}}]{Shin-transition}
\bibinfo{author}{\bibfnamefont{T.}~\bibnamefont{Shimojima}},
  \bibinfo{author}{\bibfnamefont{K.}~\bibnamefont{Ishizaka}},
  \bibinfo{author}{\bibfnamefont{Y.}~\bibnamefont{Ishida}},
  \bibinfo{author}{\bibfnamefont{N.}~\bibnamefont{Katayama}},
  \bibinfo{author}{\bibfnamefont{K.}~\bibnamefont{Ohgushi}},
  \bibinfo{author}{\bibfnamefont{T.}~\bibnamefont{Kiss}},
  \bibinfo{author}{\bibfnamefont{M.}~\bibnamefont{Okawa}},
  \bibinfo{author}{\bibfnamefont{T.}~\bibnamefont{Togashi}},
  \bibinfo{author}{\bibfnamefont{X.-Y.} \bibnamefont{Wang}},
  \bibinfo{author}{\bibfnamefont{C.-T.} \bibnamefont{Chen}},
  \bibnamefont{et~al.}, \bibinfo{journal}{Phys. Rev. Lett.}
  \textbf{\bibinfo{volume}{104}}, \bibinfo{pages}{057002}
  (\bibinfo{year}{2010}).

\bibitem[{\citenamefont{de~Jong et~al.}(2009)\citenamefont{de~Jong, Huang,
  Huisman, Massee, Thirupathaiah, Gorgoi, Schaefers, Follath, Goedkoop, and
  Golden}}]{Golden-ARPES-surface}
\bibinfo{author}{\bibfnamefont{S.}~\bibnamefont{de~Jong}},
  \bibinfo{author}{\bibfnamefont{Y.}~\bibnamefont{Huang}},
  \bibinfo{author}{\bibfnamefont{R.}~\bibnamefont{Huisman}},
  \bibinfo{author}{\bibfnamefont{F.}~\bibnamefont{Massee}},
  \bibinfo{author}{\bibfnamefont{S.}~\bibnamefont{Thirupathaiah}},
  \bibinfo{author}{\bibfnamefont{M.}~\bibnamefont{Gorgoi}},
  \bibinfo{author}{\bibfnamefont{F.}~\bibnamefont{Schaefers}},
  \bibinfo{author}{\bibfnamefont{R.}~\bibnamefont{Follath}},
  \bibinfo{author}{\bibfnamefont{J.~B.} \bibnamefont{Goedkoop}},
  \bibnamefont{and} \bibinfo{author}{\bibfnamefont{M.~S.}
  \bibnamefont{Golden}}, \bibinfo{journal}{Phys. Rev. B}
  \textbf{\bibinfo{volume}{79}}, \bibinfo{pages}{115125}
  (\bibinfo{year}{2009}).

\bibitem[{\citenamefont{Fink et~al.}(2009)\citenamefont{Fink, Thirupathaiah,
  Ovsyannikov, D\"urr, Follath, Huang, de~Jong, Golden, Zhang, Jeschke
  et~al.}}]{Fink-BaFe2As2}
\bibinfo{author}{\bibfnamefont{J.}~\bibnamefont{Fink}},
  \bibinfo{author}{\bibfnamefont{S.}~\bibnamefont{Thirupathaiah}},
  \bibinfo{author}{\bibfnamefont{R.}~\bibnamefont{Ovsyannikov}},
  \bibinfo{author}{\bibfnamefont{H.~A.} \bibnamefont{D\"urr}},
  \bibinfo{author}{\bibfnamefont{R.}~\bibnamefont{Follath}},
  \bibinfo{author}{\bibfnamefont{Y.}~\bibnamefont{Huang}},
  \bibinfo{author}{\bibfnamefont{S.}~\bibnamefont{de~Jong}},
  \bibinfo{author}{\bibfnamefont{M.~S.} \bibnamefont{Golden}},
  \bibinfo{author}{\bibfnamefont{Y.-Z.} \bibnamefont{Zhang}},
  \bibinfo{author}{\bibfnamefont{H.~O.} \bibnamefont{Jeschke}},
  \bibnamefont{et~al.}, \bibinfo{journal}{Phys. Rev. B}
  \textbf{\bibinfo{volume}{79}}, \bibinfo{pages}{155118}
  (\bibinfo{year}{2009}).

\bibitem[{\citenamefont{Malaeb et~al.}(2009)\citenamefont{Malaeb, Yoshida,
  Fujimori, Kubota, Ono, Kihou, M.~Shirage, Kito, Iyo, Eisaki
  et~al.}}]{Malaeb-3d-ARPES}
\bibinfo{author}{\bibfnamefont{W.}~\bibnamefont{Malaeb}},
  \bibinfo{author}{\bibfnamefont{T.}~\bibnamefont{Yoshida}},
  \bibinfo{author}{\bibfnamefont{A.}~\bibnamefont{Fujimori}},
  \bibinfo{author}{\bibfnamefont{M.}~\bibnamefont{Kubota}},
  \bibinfo{author}{\bibfnamefont{K.}~\bibnamefont{Ono}},
  \bibinfo{author}{\bibfnamefont{K.}~\bibnamefont{Kihou}},
  \bibinfo{author}{\bibfnamefont{P.}~\bibnamefont{M.~Shirage}},
  \bibinfo{author}{\bibfnamefont{H.}~\bibnamefont{Kito}},
  \bibinfo{author}{\bibfnamefont{A.}~\bibnamefont{Iyo}},
  \bibinfo{author}{\bibfnamefont{H.}~\bibnamefont{Eisaki}},
  \bibnamefont{et~al.}, \bibinfo{journal}{J. Phys. Soc. Jpn.}
  \textbf{\bibinfo{volume}{78}}, \bibinfo{pages}{123706}
  (\bibinfo{year}{2009}).

\bibitem[{\citenamefont{Singh}(2009)}]{singh-pnictides-bands}
\bibinfo{author}{\bibfnamefont{D.~J.} \bibnamefont{Singh}},
  \bibinfo{journal}{Physica C: Superconductivity}
  \textbf{\bibinfo{volume}{469}}, \bibinfo{pages}{418} (\bibinfo{year}{2009}).

\bibitem[{\citenamefont{Vildosola et~al.}(2008)\citenamefont{Vildosola,
  Pourovskii, Arita, Biermann, and Georges}}]{LaOFeAs-veronica}
\bibinfo{author}{\bibfnamefont{V.}~\bibnamefont{Vildosola}},
  \bibinfo{author}{\bibfnamefont{L.}~\bibnamefont{Pourovskii}},
  \bibinfo{author}{\bibfnamefont{R.}~\bibnamefont{Arita}},
  \bibinfo{author}{\bibfnamefont{S.}~\bibnamefont{Biermann}}, \bibnamefont{and}
  \bibinfo{author}{\bibfnamefont{A.}~\bibnamefont{Georges}},
  \bibinfo{journal}{Phys. Rev. B} \textbf{\bibinfo{volume}{78}},
  \bibinfo{pages}{064518} (\bibinfo{year}{2008}).

\bibitem[{\citenamefont{Mazin et~al.}(2008)\citenamefont{Mazin, Singh,
  Johannes, and Du}}]{Mazin-Order-Parameter}
\bibinfo{author}{\bibfnamefont{I.~I.} \bibnamefont{Mazin}},
  \bibinfo{author}{\bibfnamefont{D.~J.} \bibnamefont{Singh}},
  \bibinfo{author}{\bibfnamefont{M.~D.} \bibnamefont{Johannes}},
  \bibnamefont{and} \bibinfo{author}{\bibfnamefont{M.~H.} \bibnamefont{Du}},
  \bibinfo{journal}{Phys. Rev. Lett.} \textbf{\bibinfo{volume}{101}},
  \bibinfo{pages}{057003} (\bibinfo{year}{2008}).

\bibitem[{\citenamefont{Haule et~al.}(2008)\citenamefont{Haule, Shim, and
  Kotliar}}]{LaOFeAs-haule-2008}
\bibinfo{author}{\bibfnamefont{K.}~\bibnamefont{Haule}},
  \bibinfo{author}{\bibfnamefont{J.~H.} \bibnamefont{Shim}}, \bibnamefont{and}
  \bibinfo{author}{\bibfnamefont{G.}~\bibnamefont{Kotliar}},
  \bibinfo{journal}{Phys. Rev. Lett.} \textbf{\bibinfo{volume}{100}},
  \bibinfo{pages}{226402} (\bibinfo{year}{2008}).

\bibitem[{\citenamefont{Aichhorn et~al.}(2009)\citenamefont{Aichhorn,
  Pourovskii, Vildosola, Ferrero, Parcollet, Miyake, Georges, and
  Biermann}}]{cRPA-DMFT-LaOFeAs-markus}
\bibinfo{author}{\bibfnamefont{M.}~\bibnamefont{Aichhorn}},
  \bibinfo{author}{\bibfnamefont{L.}~\bibnamefont{Pourovskii}},
  \bibinfo{author}{\bibfnamefont{V.}~\bibnamefont{Vildosola}},
  \bibinfo{author}{\bibfnamefont{M.}~\bibnamefont{Ferrero}},
  \bibinfo{author}{\bibfnamefont{O.}~\bibnamefont{Parcollet}},
  \bibinfo{author}{\bibfnamefont{T.}~\bibnamefont{Miyake}},
  \bibinfo{author}{\bibfnamefont{A.}~\bibnamefont{Georges}}, \bibnamefont{and}
  \bibinfo{author}{\bibfnamefont{S.}~\bibnamefont{Biermann}},
  \bibinfo{journal}{Phys. Rev. B} \textbf{\bibinfo{volume}{80}},
  \bibinfo{pages}{085101} (\bibinfo{year}{2009}).

\bibitem[{\citenamefont{Aichhorn et~al.}(2010)\citenamefont{Aichhorn, Biermann,
  Miyake, Georges, and Imada}}]{cRPA-DMFT-FeSe-markus}
\bibinfo{author}{\bibfnamefont{M.}~\bibnamefont{Aichhorn}},
  \bibinfo{author}{\bibfnamefont{S.}~\bibnamefont{Biermann}},
  \bibinfo{author}{\bibfnamefont{T.}~\bibnamefont{Miyake}},
  \bibinfo{author}{\bibfnamefont{A.}~\bibnamefont{Georges}}, \bibnamefont{and}
  \bibinfo{author}{\bibfnamefont{M.}~\bibnamefont{Imada}},
  \bibinfo{journal}{Phys. Rev. B} \textbf{\bibinfo{volume}{82}},
  \bibinfo{pages}{064504} (\bibinfo{year}{2010}).

\bibitem[{\citenamefont{Ferber et~al.}(2012)\citenamefont{Ferber, Foyevtsova,
  Valent{\'i}, and Jeschke}}]{Valenti-LiFeAs}
\bibinfo{author}{\bibfnamefont{J.}~\bibnamefont{Ferber}},
  \bibinfo{author}{\bibfnamefont{K.}~\bibnamefont{Foyevtsova}},
  \bibinfo{author}{\bibfnamefont{R.}~\bibnamefont{Valent{\'i}}},
  \bibnamefont{and} \bibinfo{author}{\bibfnamefont{H.~O.}
  \bibnamefont{Jeschke}}, \bibinfo{journal}{Phys. Rev. B}
  \textbf{\bibinfo{volume}{85}}, \bibinfo{pages}{094505}
  (\bibinfo{year}{2012}).

\bibitem[{\citenamefont{Hansmann et~al.}(2010)\citenamefont{Hansmann, Arita,
  Toschi, Sakai, Sangiovanni, and Held}}]{LaOFeAs-hansmann-2010}
\bibinfo{author}{\bibfnamefont{P.}~\bibnamefont{Hansmann}},
  \bibinfo{author}{\bibfnamefont{R.}~\bibnamefont{Arita}},
  \bibinfo{author}{\bibfnamefont{A.}~\bibnamefont{Toschi}},
  \bibinfo{author}{\bibfnamefont{S.}~\bibnamefont{Sakai}},
  \bibinfo{author}{\bibfnamefont{G.}~\bibnamefont{Sangiovanni}},
  \bibnamefont{and} \bibinfo{author}{\bibfnamefont{K.}~\bibnamefont{Held}},
  \bibinfo{journal}{Phys. Rev. Lett.} \textbf{\bibinfo{volume}{104}},
  \bibinfo{pages}{197002} (\bibinfo{year}{2010}).

\bibitem[{\citenamefont{Anisimov et~al.}(2009)\citenamefont{Anisimov, Korotin,
  Korotin, Kozhevnikov, Kunes, Shorikov, Skornyakov, and
  Streltsov}}]{LaOFeAs-anisimov-2009}
\bibinfo{author}{\bibfnamefont{V.~I.} \bibnamefont{Anisimov}},
  \bibinfo{author}{\bibfnamefont{D.}~\bibnamefont{Korotin}},
  \bibinfo{author}{\bibfnamefont{M.}~\bibnamefont{Korotin}},
  \bibinfo{author}{\bibfnamefont{A.~V.} \bibnamefont{Kozhevnikov}},
  \bibinfo{author}{\bibfnamefont{J.}~\bibnamefont{Kunes}},
  \bibinfo{author}{\bibfnamefont{A.~O.} \bibnamefont{Shorikov}},
  \bibinfo{author}{\bibfnamefont{S.~L.} \bibnamefont{Skornyakov}},
  \bibnamefont{and} \bibinfo{author}{\bibfnamefont{S.~V.}
  \bibnamefont{Streltsov}}, \bibinfo{journal}{J. Phys. Condens. Matter}
  \textbf{\bibinfo{volume}{21}}, \bibinfo{pages}{075602}
  (\bibinfo{year}{2009}).

\bibitem[{\citenamefont{Werner et~al.}(2012)\citenamefont{Werner, Casula,
  Miyake, Aryasetiawan, Millis, and Biermann}}]{udyn-werner}
\bibinfo{author}{\bibfnamefont{P.}~\bibnamefont{Werner}},
  \bibinfo{author}{\bibfnamefont{M.}~\bibnamefont{Casula}},
  \bibinfo{author}{\bibfnamefont{T.}~\bibnamefont{Miyake}},
  \bibinfo{author}{\bibfnamefont{F.}~\bibnamefont{Aryasetiawan}},
  \bibinfo{author}{\bibfnamefont{A.~J.} \bibnamefont{Millis}},
  \bibnamefont{and} \bibinfo{author}{\bibfnamefont{S.}~\bibnamefont{Biermann}},
  \bibinfo{journal}{Nature Physics} \textbf{\bibinfo{volume}{8}},
  \bibinfo{pages}{331} (\bibinfo{year}{2012}).

\bibitem[{\citenamefont{Wang et~al.}(2010)\citenamefont{Wang, Qian, Xu, Dai,
  and Fang}}]{Fang-Gutzwiller-pnictides}
\bibinfo{author}{\bibfnamefont{G.}~\bibnamefont{Wang}},
  \bibinfo{author}{\bibfnamefont{Y.}~\bibnamefont{Qian}},
  \bibinfo{author}{\bibfnamefont{G.}~\bibnamefont{Xu}},
  \bibinfo{author}{\bibfnamefont{X.}~\bibnamefont{Dai}}, \bibnamefont{and}
  \bibinfo{author}{\bibfnamefont{Z.}~\bibnamefont{Fang}},
  \bibinfo{journal}{Phys. Rev. Lett.} \textbf{\bibinfo{volume}{104}},
  \bibinfo{pages}{047002} (\bibinfo{year}{2010}).

\bibitem[{\citenamefont{Anisimov et~al.}(1997)\citenamefont{Anisimov,
  Poteryaev, Korotin, Anokhin, and Kotliar}}]{LDA+DMFT-anisimov-1997}
\bibinfo{author}{\bibfnamefont{V.~I.} \bibnamefont{Anisimov}},
  \bibinfo{author}{\bibfnamefont{A.}~\bibnamefont{Poteryaev}},
  \bibinfo{author}{\bibfnamefont{M.}~\bibnamefont{Korotin}},
  \bibinfo{author}{\bibfnamefont{A.}~\bibnamefont{Anokhin}}, \bibnamefont{and}
  \bibinfo{author}{\bibfnamefont{G.}~\bibnamefont{Kotliar}},
  \bibinfo{journal}{J. Phys. Condens. Matter} \textbf{\bibinfo{volume}{9}},
  \bibinfo{pages}{7359} (\bibinfo{year}{1997}).

\bibitem[{\citenamefont{Lichtenstein and Katsnelson}(1998)}]{LDA+DMFT-licht}
\bibinfo{author}{\bibfnamefont{A.~I.} \bibnamefont{Lichtenstein}}
  \bibnamefont{and} \bibinfo{author}{\bibfnamefont{M.~I.}
  \bibnamefont{Katsnelson}}, \bibinfo{journal}{Phys. Rev. B}
  \textbf{\bibinfo{volume}{57}}, \bibinfo{pages}{6884} (\bibinfo{year}{1998}).

\bibitem[{\citenamefont{Kotliar and Vollhardt}(2004)}]{PT-kotliar}
\bibinfo{author}{\bibfnamefont{G.}~\bibnamefont{Kotliar}} \bibnamefont{and}
  \bibinfo{author}{\bibfnamefont{D.}~\bibnamefont{Vollhardt}},
  \bibinfo{journal}{Physics Today} \textbf{\bibinfo{volume}{57}},
  \bibinfo{pages}{53} (\bibinfo{year}{2004}).

\bibitem[{\citenamefont{Biermann}(2006)}]{biermann_ldadmft}
\bibinfo{author}{\bibfnamefont{S.}~\bibnamefont{Biermann}}, in
  \emph{\bibinfo{booktitle}{Encyclopedia of Materials: Science and
  Technology}}, edited by \bibinfo{editor}{\bibfnamefont{K.~H.~J.}
  \bibnamefont{Buschow}}, \bibinfo{editor}{\bibfnamefont{R.~W.}
  \bibnamefont{Cahn}}, \bibinfo{editor}{\bibfnamefont{M.~C.}
  \bibnamefont{Flemings}}, \bibinfo{editor}{\bibfnamefont{B.}
  \bibnamefont{Ilschner~(print)}}, \bibinfo{editor}{\bibfnamefont{E.~J.}
  \bibnamefont{Kramer}},
  \bibinfo{editor}{\bibfnamefont{S.}~\bibnamefont{Mahajan}},
  \bibnamefont{and} \bibinfo{editor}{\bibfnamefont{P.}
  \bibnamefont{Veyssi{\`e}re~(updates)}} (\bibinfo{publisher}{Elsevier},
  \bibinfo{address}{Oxford}, \bibinfo{year}{2006}), pp. \bibinfo{pages}{1 --
  9}.

\bibitem[{\citenamefont{Held et~al.}(2006)\citenamefont{Held, Nekrasov, Keller,
  Eyert, Bl{\"u}mer, McMahan, Scalettar, Pruschke, Anisimov, and
  Vollhardt}}]{held_psik}
\bibinfo{author}{\bibfnamefont{K.}~\bibnamefont{Held}},
  \bibinfo{author}{\bibfnamefont{I.~A.} \bibnamefont{Nekrasov}},
  \bibinfo{author}{\bibfnamefont{G.}~\bibnamefont{Keller}},
  \bibinfo{author}{\bibfnamefont{V.}~\bibnamefont{Eyert}},
  \bibinfo{author}{\bibfnamefont{N.}~\bibnamefont{Bl{\"u}mer}},
  \bibinfo{author}{\bibfnamefont{A.~K.} \bibnamefont{McMahan}},
  \bibinfo{author}{\bibfnamefont{R.~T.} \bibnamefont{Scalettar}},
  \bibinfo{author}{\bibfnamefont{T.}~\bibnamefont{Pruschke}},
  \bibinfo{author}{\bibfnamefont{V.~I.} \bibnamefont{Anisimov}},
  \bibnamefont{and}
  \bibinfo{author}{\bibfnamefont{D.}~\bibnamefont{Vollhardt}},
  \bibinfo{journal}{physica status solidi (b)} \textbf{\bibinfo{volume}{243}}
  (\bibinfo{year}{2006}), \bibinfo{note}{psi-k Newsletter, 56 (65) 2003}.

\bibitem[{\citenamefont{Kotliar et~al.}(2006)\citenamefont{Kotliar, Savrasov,
  Haule, Oudovenko, Parcollet, and Marianetti}}]{kotliar-review-DMFT}
\bibinfo{author}{\bibfnamefont{G.}~\bibnamefont{Kotliar}},
  \bibinfo{author}{\bibfnamefont{S.~Y.} \bibnamefont{Savrasov}},
  \bibinfo{author}{\bibfnamefont{K.}~\bibnamefont{Haule}},
  \bibinfo{author}{\bibfnamefont{V.~S.} \bibnamefont{Oudovenko}},
  \bibinfo{author}{\bibfnamefont{O.}~\bibnamefont{Parcollet}},
  \bibnamefont{and} \bibinfo{author}{\bibfnamefont{C.~A.}
  \bibnamefont{Marianetti}}, \bibinfo{journal}{Rev. Mod. Phys.}
  \textbf{\bibinfo{volume}{78}}, \bibinfo{pages}{865} (\bibinfo{year}{2006}).

\bibitem[{\citenamefont{Min\'ar et~al.}(2005)\citenamefont{Min\'ar, Chioncel,
  Perlov, Ebert, Katsnelson, and Lichtenstein}}]{LDA-DMFT-minar}
\bibinfo{author}{\bibfnamefont{J.}~\bibnamefont{Min\'ar}},
  \bibinfo{author}{\bibfnamefont{L.}~\bibnamefont{Chioncel}},
  \bibinfo{author}{\bibfnamefont{A.}~\bibnamefont{Perlov}},
  \bibinfo{author}{\bibfnamefont{H.}~\bibnamefont{Ebert}},
  \bibinfo{author}{\bibfnamefont{M.~I.} \bibnamefont{Katsnelson}},
  \bibnamefont{and} \bibinfo{author}{\bibfnamefont{A.~I.}
  \bibnamefont{Lichtenstein}}, \bibinfo{journal}{Phys. Rev. B}
  \textbf{\bibinfo{volume}{72}}, \bibinfo{pages}{045125}
  (\bibinfo{year}{2005}).

\bibitem[{\citenamefont{Tomczak
  et~al.}(2012{\natexlab{a}})\citenamefont{Tomczak, van Schilfgaarde, and
  Kotliar}}]{pnictides-QSGW-Jan}
\bibinfo{author}{\bibfnamefont{J.~M.} \bibnamefont{Tomczak}},
  \bibinfo{author}{\bibfnamefont{M.}~\bibnamefont{van Schilfgaarde}},
  \bibnamefont{and} \bibinfo{author}{\bibfnamefont{G.}~\bibnamefont{Kotliar}},
  \bibinfo{journal}{Phys. Rev. Lett.} \textbf{\bibinfo{volume}{109}},
  \bibinfo{pages}{237010} (\bibinfo{year}{2012}{\natexlab{a}}).

\bibitem[{\citenamefont{Brouet et~al.}(2013)\citenamefont{Brouet, Lin, Texier,
  Bobroff, Taleb-Ibrahimi, Le~F{\`e}vre, Bertran, Casula, Werner, Biermann
  et~al.}}]{FS-Brouet}
\bibinfo{author}{\bibfnamefont{V.}~\bibnamefont{Brouet}},
  \bibinfo{author}{\bibfnamefont{P.-H.} \bibnamefont{Lin}},
  \bibinfo{author}{\bibfnamefont{Y.}~\bibnamefont{Texier}},
  \bibinfo{author}{\bibfnamefont{J.}~\bibnamefont{Bobroff}},
  \bibinfo{author}{\bibfnamefont{A.}~\bibnamefont{Taleb-Ibrahimi}},
  \bibinfo{author}{\bibfnamefont{P.}~\bibnamefont{Le~F{\`e}vre}},
  \bibinfo{author}{\bibfnamefont{F.}~\bibnamefont{Bertran}},
  \bibinfo{author}{\bibfnamefont{M.}~\bibnamefont{Casula}},
  \bibinfo{author}{\bibfnamefont{P.}~\bibnamefont{Werner}},
  \bibinfo{author}{\bibfnamefont{S.}~\bibnamefont{Biermann}},
  \bibnamefont{et~al.}, \bibinfo{journal}{Phys. Rev. Lett.}
  \textbf{\bibinfo{volume}{110}}, \bibinfo{pages}{167002}
  (\bibinfo{year}{2013}).

\bibitem[{\citenamefont{van Schilfgaarde et~al.}(2006)\citenamefont{van
  Schilfgaarde, Kotani, and Faleev}}]{scGW-kotani}
\bibinfo{author}{\bibfnamefont{M.}~\bibnamefont{van Schilfgaarde}},
  \bibinfo{author}{\bibfnamefont{T.}~\bibnamefont{Kotani}}, \bibnamefont{and}
  \bibinfo{author}{\bibfnamefont{S.}~\bibnamefont{Faleev}},
  \bibinfo{journal}{Phys. Rev. Lett.} \textbf{\bibinfo{volume}{96}},
  \bibinfo{pages}{226402} (\bibinfo{year}{2006}).

\bibitem[{\citenamefont{Ayral et~al.}(2012)\citenamefont{Ayral, Werner, and
  Biermann}}]{Thomas-PRL-GW+DMFT}
\bibinfo{author}{\bibfnamefont{T.}~\bibnamefont{Ayral}},
  \bibinfo{author}{\bibfnamefont{P.}~\bibnamefont{Werner}}, \bibnamefont{and}
  \bibinfo{author}{\bibfnamefont{S.}~\bibnamefont{Biermann}},
  \bibinfo{journal}{Phys. Rev. Lett.} \textbf{\bibinfo{volume}{109}},
  \bibinfo{pages}{226401} (\bibinfo{year}{2012}).

\bibitem[{\citenamefont{Haule and Kotliar}(2009)}]{LaOFeAs-kotliar-2009}
\bibinfo{author}{\bibfnamefont{K.}~\bibnamefont{Haule}} \bibnamefont{and}
  \bibinfo{author}{\bibfnamefont{G.}~\bibnamefont{Kotliar}},
  \bibinfo{journal}{New J. Phys.} \textbf{\bibinfo{volume}{11}},
  \bibinfo{pages}{025021} (\bibinfo{year}{2009}).

\bibitem[{\citenamefont{Biermann et~al.}(2003)\citenamefont{Biermann,
  Aryasetiawan, and Georges}}]{GW+DMFT-biermann}
\bibinfo{author}{\bibfnamefont{S.}~\bibnamefont{Biermann}},
  \bibinfo{author}{\bibfnamefont{F.}~\bibnamefont{Aryasetiawan}},
  \bibnamefont{and} \bibinfo{author}{\bibfnamefont{A.}~\bibnamefont{Georges}},
  \bibinfo{journal}{Phys. Rev. Lett.} \textbf{\bibinfo{volume}{90}},
  \bibinfo{pages}{086402} (\bibinfo{year}{2003}).

\bibitem[{\citenamefont{Hedin}(1965)}]{Hedin-1965}
\bibinfo{author}{\bibfnamefont{L.}~\bibnamefont{Hedin}},
  \bibinfo{journal}{Phys. Rev.} \textbf{\bibinfo{volume}{139}},
  \bibinfo{pages}{A796} (\bibinfo{year}{1965}).

\bibitem[{\citenamefont{Bylander and Kleinman}(1990)}]{Bylander-hybrid}
\bibinfo{author}{\bibfnamefont{D.~M.} \bibnamefont{Bylander}} \bibnamefont{and}
  \bibinfo{author}{\bibfnamefont{L.}~\bibnamefont{Kleinman}},
  \bibinfo{journal}{Phys. Rev. B} \textbf{\bibinfo{volume}{41}},
  \bibinfo{pages}{7868} (\bibinfo{year}{1990}).

\bibitem[{\citenamefont{Seidl et~al.}(1996)\citenamefont{Seidl, G\"orling,
  Vogl, Majewski, and Levy}}]{Seidl-hybrid}
\bibinfo{author}{\bibfnamefont{A.}~\bibnamefont{Seidl}},
  \bibinfo{author}{\bibfnamefont{A.}~\bibnamefont{G\"orling}},
  \bibinfo{author}{\bibfnamefont{P.}~\bibnamefont{Vogl}},
  \bibinfo{author}{\bibfnamefont{J.~A.} \bibnamefont{Majewski}},
  \bibnamefont{and} \bibinfo{author}{\bibfnamefont{M.}~\bibnamefont{Levy}},
  \bibinfo{journal}{Phys. Rev. B} \textbf{\bibinfo{volume}{53}},
  \bibinfo{pages}{3764} (\bibinfo{year}{1996}).

\bibitem[{\citenamefont{Xu et~al.}(2013)\citenamefont{Xu, Richard, van
  Roekeghem, Zhang, Miao, Zhang, Qian, Ferrero, Sefat, Biermann
  et~al.}}]{BaCo2As2-Nan}
\bibinfo{author}{\bibfnamefont{N.}~\bibnamefont{Xu}},
  \bibinfo{author}{\bibfnamefont{P.}~\bibnamefont{Richard}},
  \bibinfo{author}{\bibfnamefont{A.}~\bibnamefont{van Roekeghem}},
  \bibinfo{author}{\bibfnamefont{P.}~\bibnamefont{Zhang}},
  \bibinfo{author}{\bibfnamefont{H.}~\bibnamefont{Miao}},
  \bibinfo{author}{\bibfnamefont{W.-L.} \bibnamefont{Zhang}},
  \bibinfo{author}{\bibfnamefont{T.}~\bibnamefont{Qian}},
  \bibinfo{author}{\bibfnamefont{M.}~\bibnamefont{Ferrero}},
  \bibinfo{author}{\bibfnamefont{A.~S.} \bibnamefont{Sefat}},
  \bibinfo{author}{\bibfnamefont{S.}~\bibnamefont{Biermann}},
  \bibnamefont{et~al.}, \bibinfo{journal}{Phys. Rev. X}
  \textbf{\bibinfo{volume}{3}}, \bibinfo{pages}{011006} (\bibinfo{year}{2013}).

\bibitem[{\citenamefont{Dhaka et~al.}(2013)\citenamefont{Dhaka, Lee, Anand,
  Johnston, Harmon, and Kaminski}}]{BaCo2As2-Dakha}
\bibinfo{author}{\bibfnamefont{R.~S.} \bibnamefont{Dhaka}},
  \bibinfo{author}{\bibfnamefont{Y.}~\bibnamefont{Lee}},
  \bibinfo{author}{\bibfnamefont{V.~K.} \bibnamefont{Anand}},
  \bibinfo{author}{\bibfnamefont{D.~C.} \bibnamefont{Johnston}},
  \bibinfo{author}{\bibfnamefont{B.~N.} \bibnamefont{Harmon}},
  \bibnamefont{and} \bibinfo{author}{\bibfnamefont{A.}~\bibnamefont{Kaminski}},
  \bibinfo{journal}{Phys. Rev. B} \textbf{\bibinfo{volume}{87}},
  \bibinfo{pages}{214516} (\bibinfo{year}{2013}).

\bibitem[{\citenamefont{Sefat et~al.}(2009)\citenamefont{Sefat, Singh, Jin,
  McGuire, Sales, and Mandrus}}]{BaCo2As2-Sefat}
\bibinfo{author}{\bibfnamefont{A.~S.} \bibnamefont{Sefat}},
  \bibinfo{author}{\bibfnamefont{D.~J.} \bibnamefont{Singh}},
  \bibinfo{author}{\bibfnamefont{R.}~\bibnamefont{Jin}},
  \bibinfo{author}{\bibfnamefont{M.~A.} \bibnamefont{McGuire}},
  \bibinfo{author}{\bibfnamefont{B.~C.} \bibnamefont{Sales}}, \bibnamefont{and}
  \bibinfo{author}{\bibfnamefont{D.}~\bibnamefont{Mandrus}},
  \bibinfo{journal}{Phys. Rev. B} \textbf{\bibinfo{volume}{79}},
  \bibinfo{pages}{024512} (\bibinfo{year}{2009}).

\bibitem[{\citenamefont{Pandey et~al.}(2013)\citenamefont{Pandey, Quirinale,
  Jayasekara, Sapkota, Kim, Dhaka, Lee, Heitmann, Stephens, Ogloblichev
  et~al.}}]{Johnston-SrCo2As2}
\bibinfo{author}{\bibfnamefont{A.}~\bibnamefont{Pandey}},
  \bibinfo{author}{\bibfnamefont{D.~G.} \bibnamefont{Quirinale}},
  \bibinfo{author}{\bibfnamefont{W.}~\bibnamefont{Jayasekara}},
  \bibinfo{author}{\bibfnamefont{A.}~\bibnamefont{Sapkota}},
  \bibinfo{author}{\bibfnamefont{M.~G.} \bibnamefont{Kim}},
  \bibinfo{author}{\bibfnamefont{R.~S.} \bibnamefont{Dhaka}},
  \bibinfo{author}{\bibfnamefont{Y.}~\bibnamefont{Lee}},
  \bibinfo{author}{\bibfnamefont{T.~W.} \bibnamefont{Heitmann}},
  \bibinfo{author}{\bibfnamefont{P.~W.} \bibnamefont{Stephens}},
  \bibinfo{author}{\bibfnamefont{V.}~\bibnamefont{Ogloblichev}},
  \bibnamefont{et~al.}, \bibinfo{journal}{Phys. Rev. B}
  \textbf{\bibinfo{volume}{88}}, \bibinfo{pages}{014526}
  (\bibinfo{year}{2013}).

\bibitem[{\citenamefont{Jayasekara et~al.}(2013)\citenamefont{Jayasekara, Lee,
  Pandey, Tucker, Sapkota, Lamsal, Calder, Abernathy, Niedziela, Harmon
  et~al.}}]{McQueeney-SrCo2As2}
\bibinfo{author}{\bibfnamefont{W.}~\bibnamefont{Jayasekara}},
  \bibinfo{author}{\bibfnamefont{Y.}~\bibnamefont{Lee}},
  \bibinfo{author}{\bibfnamefont{A.}~\bibnamefont{Pandey}},
  \bibinfo{author}{\bibfnamefont{G.~S.} \bibnamefont{Tucker}},
  \bibinfo{author}{\bibfnamefont{A.}~\bibnamefont{Sapkota}},
  \bibinfo{author}{\bibfnamefont{J.}~\bibnamefont{Lamsal}},
  \bibinfo{author}{\bibfnamefont{S.}~\bibnamefont{Calder}},
  \bibinfo{author}{\bibfnamefont{D.~L.} \bibnamefont{Abernathy}},
  \bibinfo{author}{\bibfnamefont{J.~L.} \bibnamefont{Niedziela}},
  \bibinfo{author}{\bibfnamefont{B.~N.} \bibnamefont{Harmon}},
  \bibnamefont{et~al.}, \bibinfo{journal}{Phys. Rev. Lett}
  \textbf{\bibinfo{volume}{111}}, \bibinfo{pages}{157001}
  (\bibinfo{year}{2013}).

\bibitem[{\citenamefont{Cheng et~al.}(2012)\citenamefont{Cheng, Hu, Yuan, Dong,
  Fang, Chen, Xu, Shi, Zheng, Luo et~al.}}]{NLWang-CaCo2As2}
\bibinfo{author}{\bibfnamefont{B.}~\bibnamefont{Cheng}},
  \bibinfo{author}{\bibfnamefont{B.~F.} \bibnamefont{Hu}},
  \bibinfo{author}{\bibfnamefont{R.~H.} \bibnamefont{Yuan}},
  \bibinfo{author}{\bibfnamefont{T.}~\bibnamefont{Dong}},
  \bibinfo{author}{\bibfnamefont{A.~F.} \bibnamefont{Fang}},
  \bibinfo{author}{\bibfnamefont{Z.~G.} \bibnamefont{Chen}},
  \bibinfo{author}{\bibfnamefont{G.}~\bibnamefont{Xu}},
  \bibinfo{author}{\bibfnamefont{Y.~G.} \bibnamefont{Shi}},
  \bibinfo{author}{\bibfnamefont{P.}~\bibnamefont{Zheng}},
  \bibinfo{author}{\bibfnamefont{J.~L.} \bibnamefont{Luo}},
  \bibnamefont{et~al.}, \bibinfo{journal}{Phys. Rev. B}
  \textbf{\bibinfo{volume}{85}}, \bibinfo{pages}{144426}
  (\bibinfo{year}{2012}).

\bibitem[{\citenamefont{Ying et~al.}(2012)\citenamefont{Ying, Yan, Wang, Xiang,
  Cheng, Ye, and Chen}}]{Chen-CaSrCo2As2}
\bibinfo{author}{\bibfnamefont{J.~J.} \bibnamefont{Ying}},
  \bibinfo{author}{\bibfnamefont{Y.~J.} \bibnamefont{Yan}},
  \bibinfo{author}{\bibfnamefont{A.~F.} \bibnamefont{Wang}},
  \bibinfo{author}{\bibfnamefont{Z.~J.} \bibnamefont{Xiang}},
  \bibinfo{author}{\bibfnamefont{P.}~\bibnamefont{Cheng}},
  \bibinfo{author}{\bibfnamefont{G.~J.} \bibnamefont{Ye}}, \bibnamefont{and}
  \bibinfo{author}{\bibfnamefont{X.~H.} \bibnamefont{Chen}},
  \bibinfo{journal}{Phys. Rev. B} \textbf{\bibinfo{volume}{85}},
  \bibinfo{pages}{214414} (\bibinfo{year}{2012}).

\bibitem[{\citenamefont{Anand et~al.}(2014)\citenamefont{Anand, Dhaka, Lee,
  Harmon, Kaminski, and Johnston}}]{Johnston-CaCo1.86As2}
\bibinfo{author}{\bibfnamefont{V.~K.} \bibnamefont{Anand}},
  \bibinfo{author}{\bibfnamefont{R.~S.} \bibnamefont{Dhaka}},
  \bibinfo{author}{\bibfnamefont{Y.}~\bibnamefont{Lee}},
  \bibinfo{author}{\bibfnamefont{B.~N.} \bibnamefont{Harmon}},
  \bibinfo{author}{\bibfnamefont{A.}~\bibnamefont{Kaminski}}, \bibnamefont{and}
  \bibinfo{author}{\bibfnamefont{D.~C.} \bibnamefont{Johnston}},
  \bibinfo{journal}{Phys. Rev. B} \textbf{\bibinfo{volume}{89}},
  \bibinfo{pages}{214409} (\bibinfo{year}{2014}).

\bibitem[{\citenamefont{Casula et~al.}(2012{\natexlab{a}})\citenamefont{Casula,
  Rubtsov, and Biermann}}]{udyn-michele}
\bibinfo{author}{\bibfnamefont{M.}~\bibnamefont{Casula}},
  \bibinfo{author}{\bibfnamefont{A.}~\bibnamefont{Rubtsov}}, \bibnamefont{and}
  \bibinfo{author}{\bibfnamefont{S.}~\bibnamefont{Biermann}},
  \bibinfo{journal}{Phys. Rev. B} \textbf{\bibinfo{volume}{85}},
  \bibinfo{pages}{035115} (\bibinfo{year}{2012}{\natexlab{a}}).

\bibitem[{\citenamefont{Vaugier et~al.}(2012)\citenamefont{Vaugier, Jiang, and
  Biermann}}]{vaugier-crpa}
\bibinfo{author}{\bibfnamefont{L.}~\bibnamefont{Vaugier}},
  \bibinfo{author}{\bibfnamefont{H.}~\bibnamefont{Jiang}}, \bibnamefont{and}
  \bibinfo{author}{\bibfnamefont{S.}~\bibnamefont{Biermann}},
  \bibinfo{journal}{Phys. Rev. B} \textbf{\bibinfo{volume}{86}},
  \bibinfo{pages}{165105} (\bibinfo{year}{2012}).

\bibitem[{\citenamefont{Werner et~al.}(2006)\citenamefont{Werner, Comanac, de'
  Medici, Troyer, and Millis}}]{CTQMC-werner}
\bibinfo{author}{\bibfnamefont{P.}~\bibnamefont{Werner}},
  \bibinfo{author}{\bibfnamefont{A.}~\bibnamefont{Comanac}},
  \bibinfo{author}{\bibfnamefont{L.}~\bibnamefont{de' Medici}},
  \bibinfo{author}{\bibfnamefont{M.}~\bibnamefont{Troyer}}, \bibnamefont{and}
  \bibinfo{author}{\bibfnamefont{A.~J.} \bibnamefont{Millis}},
  \bibinfo{journal}{Phys. Rev. Lett.} \textbf{\bibinfo{volume}{97}},
  \bibinfo{pages}{076405} (\bibinfo{year}{2006}).

\bibitem[{\citenamefont{Werner and Millis}(2010)}]{CTQMC-dyn-werner}
\bibinfo{author}{\bibfnamefont{P.}~\bibnamefont{Werner}} \bibnamefont{and}
  \bibinfo{author}{\bibfnamefont{A.~J.} \bibnamefont{Millis}},
  \bibinfo{journal}{Phys. Rev. Lett.} \textbf{\bibinfo{volume}{104}},
  \bibinfo{pages}{146401} (\bibinfo{year}{2010}).

\bibitem[{\citenamefont{Ferrero and Parcollet}(2011)}]{TRIQS-website}
\bibinfo{author}{\bibfnamefont{M.}~\bibnamefont{Ferrero}} \bibnamefont{and}
  \bibinfo{author}{\bibfnamefont{O.}~\bibnamefont{Parcollet}},
  \emph{\bibinfo{title}{TRIQS: A toolbox for research on interacting quantum
  systems}} (\bibinfo{year}{2011}), \url{http://ipht.cea.fr/triqs}.

\bibitem[{\citenamefont{Casula et~al.}(2012{\natexlab{b}})\citenamefont{Casula,
  Werner, Vaugier, Aryasetiawan, Miyake, Millis, and
  Biermann}}]{udyneff-michele}
\bibinfo{author}{\bibfnamefont{M.}~\bibnamefont{Casula}},
  \bibinfo{author}{\bibfnamefont{P.}~\bibnamefont{Werner}},
  \bibinfo{author}{\bibfnamefont{L.}~\bibnamefont{Vaugier}},
  \bibinfo{author}{\bibfnamefont{F.}~\bibnamefont{Aryasetiawan}},
  \bibinfo{author}{\bibfnamefont{T.}~\bibnamefont{Miyake}},
  \bibinfo{author}{\bibfnamefont{A.~J.} \bibnamefont{Millis}},
  \bibnamefont{and} \bibinfo{author}{\bibfnamefont{S.}~\bibnamefont{Biermann}},
  \bibinfo{journal}{Phys. Rev. Lett.} \textbf{\bibinfo{volume}{109}},
  \bibinfo{pages}{126408} (\bibinfo{year}{2012}{\natexlab{b}}).

\bibitem[{\citenamefont{Kotani et~al.}(2007)\citenamefont{Kotani, van
  Schilfgaarde, and Faleev}}]{SCGW-Kotani-PRB2007}
\bibinfo{author}{\bibfnamefont{T.}~\bibnamefont{Kotani}},
  \bibinfo{author}{\bibfnamefont{M.}~\bibnamefont{van Schilfgaarde}},
  \bibnamefont{and} \bibinfo{author}{\bibfnamefont{S.~V.}
  \bibnamefont{Faleev}}, \bibinfo{journal}{Phys. Rev. B}
  \textbf{\bibinfo{volume}{76}}, \bibinfo{pages}{165106}
  (\bibinfo{year}{2007}).

\bibitem[{\citenamefont{Zhang et~al.}(2011)\citenamefont{Zhang, Richard, Qian,
  Xu, Dai, and Ding}}]{Peng-curvature}
\bibinfo{author}{\bibfnamefont{P.}~\bibnamefont{Zhang}},
  \bibinfo{author}{\bibfnamefont{P.}~\bibnamefont{Richard}},
  \bibinfo{author}{\bibfnamefont{T.}~\bibnamefont{Qian}},
  \bibinfo{author}{\bibfnamefont{Y.-M.} \bibnamefont{Xu}},
  \bibinfo{author}{\bibfnamefont{X.}~\bibnamefont{Dai}}, \bibnamefont{and}
  \bibinfo{author}{\bibfnamefont{H.}~\bibnamefont{Ding}},
  \bibinfo{journal}{Rev. Sci. Instrum.} \textbf{\bibinfo{volume}{82}},
  \bibinfo{pages}{043712} (\bibinfo{year}{2011}).

\bibitem[{\citenamefont{Tomczak et~al.}(2014)\citenamefont{Tomczak, Casula,
  Miyake, and Biermann}}]{SrVO3-GW+DMFT-Jan-long}
\bibinfo{author}{\bibfnamefont{J.~M.} \bibnamefont{Tomczak}},
  \bibinfo{author}{\bibfnamefont{M.}~\bibnamefont{Casula}},
  \bibinfo{author}{\bibfnamefont{T.}~\bibnamefont{Miyake}}, \bibnamefont{and}
  \bibinfo{author}{\bibfnamefont{S.}~\bibnamefont{Biermann}},
  \bibinfo{journal}{Phys. Rev. B} \textbf{\bibinfo{volume}{90}},
  \bibinfo{pages}{165138} (\bibinfo{year}{2014}).

\bibitem[{\citenamefont{Tomczak
  et~al.}(2012{\natexlab{b}})\citenamefont{Tomczak, Casula, Miyake,
  Aryasetiawan, and Biermann}}]{SrVO3-GW+DMFT-Jan}
\bibinfo{author}{\bibfnamefont{J.~M.} \bibnamefont{Tomczak}},
  \bibinfo{author}{\bibfnamefont{M.}~\bibnamefont{Casula}},
  \bibinfo{author}{\bibfnamefont{T.}~\bibnamefont{Miyake}},
  \bibinfo{author}{\bibfnamefont{F.}~\bibnamefont{Aryasetiawan}},
  \bibnamefont{and} \bibinfo{author}{\bibfnamefont{S.}~\bibnamefont{Biermann}},
  \bibinfo{journal}{EPL} \textbf{\bibinfo{volume}{100}}, \bibinfo{pages}{67001}
  (\bibinfo{year}{2012}{\natexlab{b}}).

\end{thebibliography}

\begin{thebibliography}{13}
\expandafter\ifx\csname natexlab\endcsname\relax\def\natexlab#1{#1}\fi
\expandafter\ifx\csname bibnamefont\endcsname\relax
  \def\bibnamefont#1{#1}\fi
\expandafter\ifx\csname bibfnamefont\endcsname\relax
  \def\bibfnamefont#1{#1}\fi
\expandafter\ifx\csname citenamefont\endcsname\relax
  \def\citenamefont#1{#1}\fi
\expandafter\ifx\csname url\endcsname\relax
  \def\url#1{\texttt{#1}}\fi
\expandafter\ifx\csname urlprefix\endcsname\relax\def\urlprefix{URL }\fi
\providecommand{\bibinfo}[2]{#2}
\providecommand{\eprint}[2][]{\url{#2}}

\bibitem[{\citenamefont{Tomczak et~al.}(2012)\citenamefont{Tomczak, van
  Schilfgaarde, and Kotliar}}]{pnictides-QSGW-Jan-s}
\bibinfo{author}{\bibfnamefont{J.~M.} \bibnamefont{Tomczak}},
  \bibinfo{author}{\bibfnamefont{M.}~\bibnamefont{van Schilfgaarde}},
  \bibnamefont{and} \bibinfo{author}{\bibfnamefont{G.}~\bibnamefont{Kotliar}},
  \bibinfo{journal}{Phys. Rev. Lett.} \textbf{\bibinfo{volume}{109}},
  \bibinfo{pages}{237010} (\bibinfo{year}{2012}).

\bibitem[{\citenamefont{Tomczak et~al.}(2014)\citenamefont{Tomczak, Casula,
  Miyake, and Biermann}}]{SrVO3-GW+DMFT-Jan-long-s}
\bibinfo{author}{\bibfnamefont{J.~M.} \bibnamefont{Tomczak}},
  \bibinfo{author}{\bibfnamefont{M.}~\bibnamefont{Casula}},
  \bibinfo{author}{\bibfnamefont{T.}~\bibnamefont{Miyake}}, \bibnamefont{and}
  \bibinfo{author}{\bibfnamefont{S.}~\bibnamefont{Biermann}},
  \bibinfo{journal}{Phys. Rev. B} \textbf{\bibinfo{volume}{90}},
  \bibinfo{pages}{165138} (\bibinfo{year}{2014}).

\bibitem[{\citenamefont{Werner and Millis}(2010)}]{CTQMC-dyn-werner-s}
\bibinfo{author}{\bibfnamefont{P.}~\bibnamefont{Werner}} \bibnamefont{and}
  \bibinfo{author}{\bibfnamefont{A.~J.} \bibnamefont{Millis}},
  \bibinfo{journal}{Phys. Rev. Lett.} \textbf{\bibinfo{volume}{104}},
  \bibinfo{pages}{146401} (\bibinfo{year}{2010}).

\bibitem[{\citenamefont{Casula et~al.}(2012)\citenamefont{Casula, Werner,
  Vaugier, Aryasetiawan, Miyake, Millis, and Biermann}}]{udyneff-michele-s}
\bibinfo{author}{\bibfnamefont{M.}~\bibnamefont{Casula}},
  \bibinfo{author}{\bibfnamefont{P.}~\bibnamefont{Werner}},
  \bibinfo{author}{\bibfnamefont{L.}~\bibnamefont{Vaugier}},
  \bibinfo{author}{\bibfnamefont{F.}~\bibnamefont{Aryasetiawan}},
  \bibinfo{author}{\bibfnamefont{T.}~\bibnamefont{Miyake}},
  \bibinfo{author}{\bibfnamefont{A.~J.} \bibnamefont{Millis}},
  \bibnamefont{and} \bibinfo{author}{\bibfnamefont{S.}~\bibnamefont{Biermann}},
  \bibinfo{journal}{Phys. Rev. Lett.} \textbf{\bibinfo{volume}{109}},
  \bibinfo{pages}{126408} (\bibinfo{year}{2012}).

\bibitem[{\citenamefont{Blaha et~al.}(2001)\citenamefont{Blaha, Schwarz,
  Madsen, Kvasnicka, and Luitz}}]{blaha_wien2k-s}
\bibinfo{author}{\bibfnamefont{P.}~\bibnamefont{Blaha}},
  \bibinfo{author}{\bibfnamefont{K.}~\bibnamefont{Schwarz}},
  \bibinfo{author}{\bibfnamefont{G.}~\bibnamefont{Madsen}},
  \bibinfo{author}{\bibfnamefont{D.}~\bibnamefont{Kvasnicka}},
  \bibnamefont{and} \bibinfo{author}{\bibfnamefont{J.}~\bibnamefont{Luitz}},
  \emph{\bibinfo{title}{\textsf{Wien2k}, {A}n {A}ugmented {P}lane
  {W}ave+{L}ocal {O}rbitals {P}rogram for {C}alculating {C}rystal
  {P}roperties}} (\bibinfo{publisher}{Tech. Universit\"at Wien, Austria},
  \bibinfo{year}{2001}).

\bibitem[{\citenamefont{Tran and Blaha}(2011)}]{wien-hf-Tran-s}
\bibinfo{author}{\bibfnamefont{F.~O.} \bibnamefont{Tran}} \bibnamefont{and}
  \bibinfo{author}{\bibfnamefont{P.}~\bibnamefont{Blaha}},
  \bibinfo{journal}{Phys. Rev. B} \textbf{\bibinfo{volume}{83}},
  \bibinfo{pages}{235118} (\bibinfo{year}{2011}).

\bibitem[{\citenamefont{Vaugier et~al.}(2012)\citenamefont{Vaugier, Jiang, and
  Biermann}}]{vaugier-crpa-s}
\bibinfo{author}{\bibfnamefont{L.}~\bibnamefont{Vaugier}},
  \bibinfo{author}{\bibfnamefont{H.}~\bibnamefont{Jiang}}, \bibnamefont{and}
  \bibinfo{author}{\bibfnamefont{S.}~\bibnamefont{Biermann}},
  \bibinfo{journal}{Phys. Rev. B} \textbf{\bibinfo{volume}{86}},
  \bibinfo{pages}{165105} (\bibinfo{year}{2012}).

\bibitem[{\citenamefont{Vaugier}(2011)}]{loig-tesis-s}
\bibinfo{author}{\bibfnamefont{L.}~\bibnamefont{Vaugier}}, Ph.D. thesis,
  \bibinfo{school}{Ecole Polytechnique, France} (\bibinfo{year}{2011}).

\bibitem[{\citenamefont{Werner et~al.}(2012)\citenamefont{Werner, Casula,
  Miyake, Aryasetiawan, Millis, and Biermann}}]{udyn-werner-s}
\bibinfo{author}{\bibfnamefont{P.}~\bibnamefont{Werner}},
  \bibinfo{author}{\bibfnamefont{M.}~\bibnamefont{Casula}},
  \bibinfo{author}{\bibfnamefont{T.}~\bibnamefont{Miyake}},
  \bibinfo{author}{\bibfnamefont{F.}~\bibnamefont{Aryasetiawan}},
  \bibinfo{author}{\bibfnamefont{A.~J.} \bibnamefont{Millis}},
  \bibnamefont{and} \bibinfo{author}{\bibfnamefont{S.}~\bibnamefont{Biermann}},
  \bibinfo{journal}{Nature Physics} \textbf{\bibinfo{volume}{8}},
  \bibinfo{pages}{331} (\bibinfo{year}{2012}).

\bibitem[{\citenamefont{Anisimov et~al.}(1991)\citenamefont{Anisimov, Zaanen,
  and Andersen}}]{cLDA-anisimov-1991-s}
\bibinfo{author}{\bibfnamefont{V.~I.} \bibnamefont{Anisimov}},
  \bibinfo{author}{\bibfnamefont{J.}~\bibnamefont{Zaanen}}, \bibnamefont{and}
  \bibinfo{author}{\bibfnamefont{O.~K.} \bibnamefont{Andersen}},
  \bibinfo{journal}{Phys. Rev. B} \textbf{\bibinfo{volume}{44}},
  \bibinfo{pages}{943} (\bibinfo{year}{1991}).

\bibitem[{\citenamefont{Biermann et~al.}(2003)\citenamefont{Biermann,
  Aryasetiawan, and Georges}}]{GW+DMFT-biermann-s}
\bibinfo{author}{\bibfnamefont{S.}~\bibnamefont{Biermann}},
  \bibinfo{author}{\bibfnamefont{F.}~\bibnamefont{Aryasetiawan}},
  \bibnamefont{and} \bibinfo{author}{\bibfnamefont{A.}~\bibnamefont{Georges}},
  \bibinfo{journal}{Phys. Rev. Lett.} \textbf{\bibinfo{volume}{90}},
  \bibinfo{pages}{086402} (\bibinfo{year}{2003}).

\bibitem[{\citenamefont{Hedin}(1999)}]{Hedin-review-s}
\bibinfo{author}{\bibfnamefont{L.}~\bibnamefont{Hedin}}, \bibinfo{journal}{J.
  Phys. Condens. Matter} \textbf{\bibinfo{volume}{11}}, \bibinfo{pages}{R489}
  (\bibinfo{year}{1999}).

\bibitem[{\citenamefont{Morikawa et~al.}(1995)\citenamefont{Morikawa, Mizokawa,
  Kobayashi, Fujimori, Eisaki, Uchida, Iga, and Nishihara}}]{SrVO3-morikawa-s}
\bibinfo{author}{\bibfnamefont{K.}~\bibnamefont{Morikawa}},
  \bibinfo{author}{\bibfnamefont{T.}~\bibnamefont{Mizokawa}},
  \bibinfo{author}{\bibfnamefont{K.}~\bibnamefont{Kobayashi}},
  \bibinfo{author}{\bibfnamefont{A.}~\bibnamefont{Fujimori}},
  \bibinfo{author}{\bibfnamefont{H.}~\bibnamefont{Eisaki}},
  \bibinfo{author}{\bibfnamefont{S.}~\bibnamefont{Uchida}},
  \bibinfo{author}{\bibfnamefont{F.}~\bibnamefont{Iga}}, \bibnamefont{and}
  \bibinfo{author}{\bibfnamefont{Y.}~\bibnamefont{Nishihara}},
  \bibinfo{journal}{Phys. Rev. B} \textbf{\bibinfo{volume}{52}},
  \bibinfo{pages}{13711} (\bibinfo{year}{1995}).

\bibitem[{\citenamefont{van Roekeghem and Biermann}(2014)}]{SrVO3-Ambroise-s}
\bibinfo{author}{\bibfnamefont{A.} \bibnamefont{van Roekeghem}} \bibnamefont{and}
  \bibinfo{author}{\bibfnamefont{S.}~\bibnamefont{Biermann}},
  \bibinfo{journal}{EPL} \textbf{\bibinfo{volume}{108}},
  \bibinfo{pages}{57003} (\bibinfo{year}{2014}).

\end{thebibliography}

\end{document}